\documentclass[aps,twocolumn,showpacs,preprintnumbers,amsmath,floatfix]{revtex4-1}

\usepackage{graphicx}
\usepackage{epsfig}
\usepackage{dcolumn}
\usepackage{bbm}
\usepackage{verbatim}
\usepackage[colorlinks,linkcolor=red,anchorcolor=green,citecolor=blue,CJKbookmarks=True]{hyperref}
\usepackage{xcolor}

\begin{document}
\title{Annihilation diagram contribution to charmonium masses}
\author{\small
	Renqiang Zhang,${}^{1,2}$ Wei Sun,${}^{1}$, Feiyu Chen,${}^{1,2}$ Ying Chen,${}^{1,2}$\footnote{cheny@ihep.ac.cn},
	Ming Gong,${}^{1,2}$ Xiangyu Jiang ${}^{1,2}$ and Zhaofeng Liu${}^{1,2}$}

\affiliation{ \small
	$^1$~Institute of High Energy Physics, Chinese Academy
	of Sciences, Beijing 100049, P.R. China \\
	$^2$~School of Physics, University of Chinese Academy
	of Sciences, Beijing 100049, P.R. China}

\begin{abstract}
In this work, we generate gauge configurations with $N_f=2$ dynamical charm quarks on anisotropic lattices. The mass shift of $1S$ and $1P$ charmonia owing to the charm quark annihilation effect can be investigated directly in a manner of unitary theory. The distillation method is adopted to treat the charm quark annihilation diagrams at a very precise level. For $1S$ charmonia, the charm quark annihilation effect almost does not change the $J/\psi$ mass, but lifts the $\eta_c$ mass by approximately 3-4 MeV. For $1P$ charmonia, this effect results in positive mass shifts of approximately 1 MeV for $\chi_{c1}$ and $h_c$, but decreases the $\chi_{c2}$ mass by approximately 3 MeV. We have not obtain a reliable result for the mass shift of $\chi_{c0}$. In addition, it is observed that the spin averaged mass of the spin-triplet $1P$ charmonia is in a good agreement with the $h_c$, as expected by the non-relativistic quark model and measured by experiments.
\\
\end{abstract}
\maketitle
\section{Introduction}

The hyperfine splitting of charmonia, namely $\Delta M_\mathrm{hyp}=M_{J/\psi}-M_{\eta_c}$ is a good quantity to test the precision of lattice QCD studies when charm quarks are involved. There have been many lattice efforts on $\Delta M_\mathrm{hyp}$ from both the quenched approximation and full-QCD simulation with dynamical quarks. A recent full-QCD calculation gives $\Delta M_\mathrm{hyp}=116.2(3.6)$ MeV 
after the continuum extrapolation~\cite{DeTar:2018uko}, which is consistent with the experimental value $\Delta M_\mathrm{hyp}^\mathrm{exp.}=113.0(0.5)$ MeV~\cite{Zyla:2020zbs}. The latest calculation carried out by the HPQCD collaboration finds $\Delta M_\mathrm{hyp}=120.3(1.1)$ MeV at the physical point after considering the quenched QED effects~\cite{Hatton:2020qhk}. Obviously, this result, with a much smaller error, still deviates from the experimental value about  $+7.3(1.2)$ MeV. Actually, most of the lattice QCD calculations of $\Delta M_\mathrm{hyp}$ do not consider the contribution of charm quark annihilation diagrams (or disconnected diagrams) to the charmonium correlation functions, which may be the major cause of the systematic uncertainty. Although this kind of contribution is expected to be highly suppressed due to the Okubo-Zweig-Iizuka rule (OZI rule)~\cite{Okubo:1963fa,Mandula:1970wz,tHooft:1976rip}, its contribution to $\Delta M_\mathrm{hyp}$ can be sizable. To be specific, the contribution of disconnected diagrams 
to the $J/\psi$ correlation function is of order $O(\alpha_s^3)$ based on a naive power 
counting of the strong coupling constant $\alpha_s$, while it is of order $O(\alpha_s^2)$ for the case of $\eta_c$. Especially, since $\eta_c$ is a flavor singlet pseudoscalar, its coupling to gluons can be enhanced due to the $U_A(1)$ anomaly of QCD, which may shift the $\eta_c$ upward in the similar sense of the origin of $\eta'$ mass~\cite{Schafer:1996wv}. Furthermore, given the lattice prediction of the pseudoscalar
glueball mass $M_G\sim 2.56$ GeV~\cite{Morningstar:1999rf,Chen:2005mg,Sun:2017ipk}, which is close to the $\eta_c$ mass, the $\eta_c$-glueball mixing may introduce an additional mass shift of $M_{\eta_c}$. Of course the light hadronic intermediate states also contribute in the presence of the light dynamical quarks. Therefore, the above effects should be taken into account if one wants to obtain a more precise theoretical determination of $\Delta M_\mathrm{hyp}$. 

There are also a few lattice studies on the annihilation diagram contribution to the masses of $J/\psi$ and $\eta_c$~\cite{McNeile:2004wu,deForcrand:2004ia}. It is found the mass of $J/\psi$ is affected little by this effect, but the $\eta_c$ mass can be sizably changed. Thus it is very possible that the annihilation diagram correction to $\Delta M_\mathrm{hyp}$ comes mainly from the shift of $M_{\eta_c}$. However, no quantitative results have been obtained due to the large statistical errors yet. A more sophisticated lattice study can be found in Ref.~\cite{Levkova:2010ft}, where the disconnected part of the pseudoscalar correlation function $D(r)$ with respect to the four-dimensional distance $r$ is complicatedly parameterized by intermediate states and quite a few excited states. The major results are that the disconnected diagrams result in an approximate +2 MeV shift of $M_{\eta_c}$ in the quenched approximation and an approximate +8 MeV shift in the $N_f=2+1$ full QCD case. However, for the lack of sea charm quarks, the mass shift of charmionium states stemming from the disconnected diagrams can only be derived indirectly and the complicated parameterization of $D(r)$ may introduce theoretical uncertainties. 
 
In this work, we investigate directly the contribution of charm sea quark loops to the masses of charmonium states. To do so, gauge configurations with charm sea quarks are necessary, such that the theory is in a unitary manner for charm quarks. As an exploratory study, we generate gauge ensembles on anisotropic lattices with $N_f=2$ degenerate sea quarks whose mass parameters are tuned to be close to the charm quark mass (The effect of light u,d and s sea quarks is ignored temporarily for simplicity). In the presence of charm sea quarks, the full correlation functions of charmonia (including the connected and disconnected parts) have well-defined spectral expressions, from which the masses can be derived directly. The key task is to perform a precise calculation on the disconnected diagrams involved. Technically, we adopt the distillation method to treat the charm quark annihilation effects~\cite{Peardon:2009gh}.
On the other hand, a large statistics is mandatory to get the precise signals of disconnected diagrams. Therefore, for the computational expense to be affordable, we generate large gauge ensembles on 
anisotropic lattices with the spatial lattice spacing $a_s$ being larger than the temporal 
lattice spacing $a_t$. Based on this numerical prescription, we would like to investigate 
directly the effect of charm quark loops on the masses of $1S$ and $1P$ charmonium states. 

As we will have a large statistics for charmonium correlation functions, we can test the 'Center of Gravity' relation for both 1P and 2P states and try to give an estimate of the mass of $h_c(2P)$. In the quark model picture, the 'Center of Gravity' mass is expected to be equal to the mass of the spin singlet~\cite{Lucha:1991vn} for charmonium states. With a large statistics, it may be possible to test this relation precisely on the lattice.

This paper is organized as follows: In Sec.~\ref{sec:numerical} we will give a brief introduction to the lattice setups for the calculation and the distillation method. Section~\ref{sec:disc} is devoted to the lattice derivation of mass shifts of $1S$ and $1P$ charmonium states owing to the disconnected diagrams. As a byproduct, In Sec.~\ref{sec:COG} we check the agreement (or deviation) between the 
'center of gravity' mass and the mass of the spin-singlet state for $1P$ and $2P$ charmonium
from the connected part of charmonium correlation functions. The summary 
can be found in Sec.~\ref{sec:summary}.

\section{Lattice setup and the distillation method}\label{sec:numerical} 
\subsection{Lattice setup}
The gauge configuration ensembles are generated on anisotropic lattices. The gluonic action is chosen to be the tadpole improved version~\cite{Morningstar:1997ff,Chen:2005mg,Sun:2017ipk,Edwards:2008ja}
\begin{eqnarray}
S_g&=&-\beta\sum\limits_{s<s'}\left[\frac{5}{9}\frac{\mathrm{Tr}P_{ss'}}{\gamma_g u_s^4}-\frac{1}{36}\frac{\mathrm{Tr}R_{ss'}}{\gamma_g u_s^6}-\frac{1}{36}\frac{\mathrm{Tr}R_{s's}}{\gamma_g u_s^6}\right]\nonumber\\
&&-\beta\sum\limits_i\left[\frac{4}{9}\frac{\gamma_g \mathrm{Tr}P_{st}}{u_s^2}-\frac{1}{36}\frac{\gamma_g
\mathrm{Tr}R_{st}}{u_s^4}\right]
\end{eqnarray}
where $P_{\mu\nu}$ and $R_{\mu\nu}$ are the $1\times 1$ plaquette variable and the $2\times 1$ rectangular Wilson loop in the $\mu\nu$ plane of the lattice, respectively, with $s,s'$ referring to the spatial direction and $\nu=t$ referring to the temporal direction. The tadpole parameter $u_s$ is defined through the spatial plaquette $u_s=\langle \frac{1}{3}\mathrm{Re}\mathrm{Tr}P_{ij}\rangle^{1/4}$, and is tuned self-consistently (The tadpole parameter of the temporal gauge links is set to be $u_t=1$ as usual). The parameter $\gamma_g$ is the bare aspect ratio parameter $\gamma_g\sim a_s/a_t$ with $a_s$ and $a_t$ being the lattice spacing in the spatial and temporal direction, respectively. For fermions, say, charm quarks in this work, the action is chosen to be the anisotropic version of tadpole improved tree-level clover action~\cite{Edwards:2008ja,HadronSpectrum:2008xlg,Sun:2017ipk} proposed by the Hadron Spectrum Collaboration
\begin{eqnarray}\label{eq:fermion}
S_f&=&\sum\limits_{x}\bar{\psi}(x)\left[m_0+\gamma_t\hat{W}_t+\sum\limits_{s}\frac{1}{\gamma_f}\gamma_s\hat{W}_s\right.\nonumber\\
&&-\frac{1}{4u_s^2}\left(\frac{\gamma_g}{\gamma_f}+\frac{1}{\xi}\right)\sum\limits_{s}\sigma_{ts}\hat{F}_{ts}\nonumber\\
&&+\left.\frac{1}{u_s^3}\frac{1}{\gamma_f}\sum\limits_{s<s'}\sigma_{ss'}\hat{F}_{ss'}\right]\psi(x)
\end{eqnarray} 
where $\hat{F}_{\mu\nu}=\frac{1}{4}\mathrm{Im}(P_{\mu\nu }(x))$ and the dimensionless Wilson operator reads
\begin{eqnarray}
\hat{W}_\mu&=&\nabla_{\mu}-\frac{1}{2}\gamma_\mu \Delta_{\mu}\nonumber\\
\nabla_{\mu}(x)&=&\frac{1}{2}\left[U_\mu (x)\delta_{x,x+\hat{\mu}}-U_\mu^\dagger(x-\hat{\mu})\delta_{x,x+\hat{\mu}}\right]\nonumber\\
\Delta_{\mu}(x)&=&U_\mu(x)\delta_{x,x+\hat{\mu}}+U_\mu^\dagger(x-\hat{\mu})\delta_{x,x-\hat{\mu}}-2.
\end{eqnarray}
For the given $\beta$ and the bare quark mass parameter $m_0$, the bare aspect ratio $\gamma_f$ for fermions  in Eq.~(\ref{eq:fermion}) is tuned along with $\gamma_g$ to give the physical aspect ratio $\xi=5$ that is derived from the dispersion relation of the pseudoscalar ($\eta_c$) meson
\begin{equation}
\sinh^2 \frac{\hat{E}}{2} =\sinh^2 \frac{\hat{m}}{2}+\frac{1}{\xi^2} \sum\limits_i \sin^2 \frac{\hat{p}_i}{2}
\end{equation}
where $\hat{E}$ and $\hat{m}$ are the energy and the mass of the hadron in lattice units, $\hat{p}_i$ is the lattice momentum $\hat{p}_i=2\pi n_i/L_s$. 

In practice, the procedure of the tuning of parameters is very complicated, since the parameters $\{\beta, \gamma_g,\gamma_f,u_s,m_0\}$ are highly correlated with each other. We omit the details here and just present the final ensemble parameters in Table~\ref{tab:ensemble}. 
\begin{table}[t]
	\centering \caption{\label{tab:ensemble}The parameters of the gauge ensembles used in this work. The spatial lattice spacing $a_s$ is set from the static potential and the Sommer's parameter $r_0=0.491$ fm. We use two bare quark mass parameter $m_0$, which give the spin average masses $M_\mathrm{avg}$ of $\eta_c$ and $J/\psi$ close to the experimental values.}
	\begin{ruledtabular}
		\begin{tabular}{cccccccc}
		    Label	&   $\beta$    &  $L_s^3\times T$   & $a_s$ (fm) & $\xi$  & $m_0$ & $N_\mathrm{cfg}$           & $M_\mathrm{avg}$ (GeV) \\
			E1      &   $2.8$    &  $16^3\times 128$ & $0.1026$ &  $5$  &  0.0408 & 6000 & $2.672$  \\
			E2        &   $2.8$    &  $16^3\times 128$ & $0.1026$ &  $5$  & 0.0549 & 6000 & $3.006$  \\
		\end{tabular}
	\end{ruledtabular}
\end{table}

 As an exploratory study, we ignore the effect of light quarks and generate gauge configurations with $N_f=2$ flavors of degenerate charm sea quarks on an $L^3\times T=16^3\times 128$ anisotropic lattice with the aspect ratio being set to be $\xi=a_s/a_t=5$. The lattice spacing $a_s$ is determined to be $a_s=0.1026$ fm through the static potential and $r_0=0.491~\mathrm{fm}$. The bare charm quark mass parameter is set from the spin average of physical $J/\psi$ and $\eta_c$ masses. This lattice setup is based on two considerations: First, it is known that the signals of gluonic operators are very noisy and demand a fairly large statistics. Secondly, we will use the distillation method to tackle with the quark annihilation diagrams (see below).  
\begin{figure}[t!]
	\includegraphics[height=6cm]{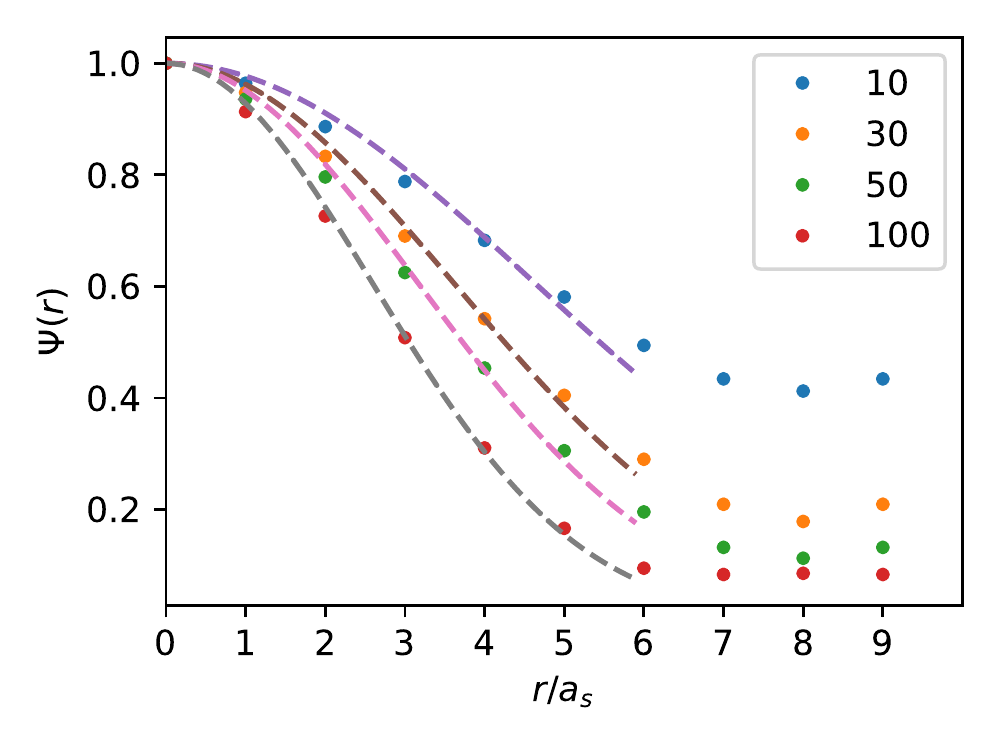}
	\caption{\label{fig:gaussian} The profiles of the smearing function $\Psi(r)$ of quark fields. The data points are the values of $\Psi(r)$ with the number of the eigenvectors of the spatial Laplacian $N=10, 30, 50$ and 100. The dashed curves are naive fits using Gaussian function forms in the interval $r/a_s\in [0,6]$. We take $N=50$ in this study.	}
\end{figure}
\subsection{Distillation method}\label{sec:distillation}
Quark-antiquark mesons can be accessed by quark bilinear operators $\bar{\psi}(\mathbf{x},t)K_U(\mathbf{x,y};t) \psi'(\mathbf{y},t)$ on the lattice. Theoretically, these operators can couple with all the meson states and possible multi-hadron systems which have the same quantum number. If one is interested in the lowest-lying states, the quark field $\psi(x)$ can be spatially smeared as 
\begin{equation}
\hat{\psi}_\alpha^a(\mathbf{x},t)=\Phi^{ab}(\mathbf{x},\mathbf{y})\psi_\alpha^b(\mathbf{y},t)
\end{equation}
where $a,b$ are color indices, $\alpha$ refers to the Dirac spin component, and $\Phi^{ab}(\mathbf{x},\mathbf{y})$ is the smearing function at timeslice $t$. Hadronic operators constructed out of smeared quark field $\hat{\psi}$ can dramatically reduce their coupling with much higher states. In order to obtain $\Phi^{ab}$, we adopt the state-of-arts approach, namely, the distillation method~\cite{Peardon:2009gh}, which is briefed as follows. 

The distillation method starts with solving the eigenvalue problem of the gauge covariant second order 
spatial Laplacian operator on each timeslice of each gauge configuration
\begin{equation}
-\nabla_{\mathbf{xy}}^2(t) v_\mathbf{y}^{(n)}(t)=\lambda_n v_\mathbf{x}^{(n)}(t)
\end{equation}
where the spatial Laplacian is expressed explicitly as
\begin{eqnarray}
-\nabla_{\mathbf{xy}}^2(t)&=&6\delta_{\mathbf{xy}}\nonumber\\
&-&\sum\limits_{j=1}^3
\left( \tilde{U}_j(\mathbf{x},t)\delta_{\mathbf{x+j,y}}+\tilde{U}_j^\dagger(\mathbf{x-j},t)\delta_{\mathbf{x-j,y}} \right)
\end{eqnarray}
where $i,j$ are spatial indices and $\tilde{U}$ is the stout-link smeared gauge link. After that, for a given Dirac index $\alpha$, each eigenvector $v_\mathbf{y}^{(n)}(t)$ is used as the source vector to solve the linear equation array
\begin{equation}
M_{\beta\delta}(\mathbf{z},t'';\mathbf{x},t')S_{\delta\alpha}(\mathbf{x},t';\mathbf{y},t)v_\mathbf{y}^{(n)}(t)=v_\mathbf{z}^{(n)}(t'')\delta_{\beta\alpha}
\end{equation} 
where the matrix $M$ is the Dirac matrix of fermions on the lattice and $S=M^{-1}$ is the corresponding fermion propagator. This procedure is repeated for all the Dirac indices $\alpha=1,2,3,4$, the timeslices $t\in [0,T-1]$ and the number of the eigenvectors $N$. Finally, multiplying the eigenvectors $v^{(n'),\dagger}_\mathbf{x}(t')$ to each of the solution vectors $S_{\delta\alpha}(\mathbf{x},t';\mathbf{y},t)v_\mathbf{y}^{(n)}(t)$ one gets for given $t',t$ and $\delta,\alpha$ the matrix elements in the subspace expanded by the $N$ eigenvectors,
\begin{equation}
\tau_{\delta\alpha}^{n'n}(t',t)=v_\mathbf{x}^{(n'),\dagger}(t')S_{\delta\alpha}(\mathbf{x},t';\mathbf{y},t)v_\mathbf{y}^{(n)}(t),
\end{equation}
which are usually called perambulators. Based on these perambulators, we can define a propagating matrix 
\begin{eqnarray}
P_{\alpha\beta}(\mathbf{x},t;\mathbf{y},t')&=& \sum\limits_{n,n'=1}^N v_\mathbf{x}^{(n)}(t)\tau_{\alpha\beta}^{nn'}(t,t')v_\mathbf{y}^{(n'),\dagger}(t')\nonumber\\
&\equiv&\Phi(\mathbf{x},\mathbf{u};t)S_{\alpha\beta}(\mathbf{u},t;\mathbf{v},t')\Phi(\mathbf{v},\mathbf{y};t')
\end{eqnarray}  
with 
\begin{equation}
\Phi(\mathbf{x,y};t)=\sum\limits_{n=1}^N v_\mathbf{x}^{(n)}(t)v_\mathbf{y}^{(n),\dagger}(t)\equiv \left[V(t)V^\dagger(t)\right]_{\mathbf{xy}}
\end{equation}
being the desired smearing function, where $V$ is a $3V_3\times N$ matrix with each column being one of the eigenvectors $v^{(n)}(t)$ for $n=1,2,\ldots, N$ and the row number $3V_3$ referring to all the spatial points and color indices. Obviously the propagating matrix $P_{\alpha\beta}(\mathbf{x},t;\mathbf{y},t')$ can be viewed as the propagator of the smeared quark field $\hat{\psi}_\alpha(\mathbf{x},t)=\Phi(\mathbf{x,y},t)\psi_\alpha(\mathbf{y},t)$ in the background gauge filed $\{U_\mu(x)\}$, namely,
\begin{equation}
P_{\alpha\beta}(\mathbf{x},t;\mathbf{y},t')=\langle \hat{\psi}_\alpha(\mathbf{x},t) \bar{\hat{\psi}}_\beta(\mathbf{y},t')\rangle_{U}.
\end{equation} 
Note that if $N=3V_3$, then the completeness of $v^{(n)}(t)$ implies that $\Phi(\mathbf{x,y};t)=\delta_\mathbf{xy}\otimes I_{\mathrm{color}}$, such that $P_{\alpha\beta}(x,y)$ is actually the all-to-all propagator of the original fermion field $\psi$. 
One can define the norm of the smearing function 
\begin{equation}
\Psi(|\mathbf{x-y}|)=\sum\limits_t\left[\mathrm{Tr}\Phi(\mathbf{x,y};t)\Phi(\mathbf{y,x};t)\right]^{1/2}
\end{equation}
to measure the degree to which the quark field is smeared. Due to the translational and the rotational
invariance, $\Psi(r)$ can be averaged over all the coordinates $\mathbf{x,y}$ that satisfying $r=|\mathbf{x-y}|$. It is easy to see that, if all the eigenvectors are involved in defining $\Phi(\mathbf{x,y};t)$, then the completeness and the orthogonalization require $\Psi(r)\propto \delta(r)$. If the number of eigenvectors involved is truncated to be $N<3V_3$, then $\Psi(r)$ is smeared from a delta function to a function falling off rapidly with respect to $r$. Figure \ref{fig:gaussian} shows the profiles of $\Psi(r)$ in data points at $N=10, 30, 50$ and 100, where $\Psi(r)$ damps faster when $N$ becomes bigger, and the dashed curves are naive fits using Gaussian function forms in the interval $r/a_s\in [0,6]$. We take $N=50$ in this study.

Now consider meson operators constructed out of the smeared quark field,
\begin{eqnarray}
\mathcal{O}_A(t)&=&\bar{\hat{\psi}}(\mathbf{x},t)K_U^A(\mathbf{x,y};t) \hat{\psi}(\mathbf{y},t),\nonumber\\ \mathcal{O}_B(t)&=&\bar{\hat{\psi}}(\mathbf{x},t)K_U^B(\mathbf{x,y};t) \hat{\psi}(\mathbf{y},t),
\end{eqnarray} 
which have the same quantum number $J^{PC}$.
The connected part of their correlation function can be expressed as 
\begin{widetext}
\begin{eqnarray}\label{eq:two-point}
C_{AB}(t,t')&=&-\mathrm{Tr}\left[ P(\mathbf{x},t;\mathbf{y'},t')K_U^B(\mathbf{y',y};t')P(\mathbf{y},t';\mathbf{x'},t)K_U^A(\mathbf{x',x};t) \right]\nonumber\\
&=& -\mathrm{Tr}\left[\Phi(\mathbf{x,u};t)S(\mathbf{u},t;\mathbf{v},t')\Phi(\mathbf{v,y'};t')K_U^B(\mathbf{y',y};t') \Phi(\mathbf{y,w};t')S(\mathbf{w},t';\mathbf{z},t)\Phi(\mathbf{z,x'};t)K_U^A(\mathbf{x',x};t)\right]\nonumber\\
&=&-\mathrm{Tr}\left[V^\dagger(\mathbf{x'},t)K_U^A(\mathbf{x'},\mathbf{x},t) V(\mathbf{x},t) \tau(t,t') V^\dagger(\mathbf{y'},t')K_U^B(\mathbf{y'},\mathbf{y},t')V(\mathbf{y},t')\tau(t',t) \right]\nonumber\\
&\equiv& -\mathrm{Tr}\left[\phi^A(t)\tau(t,t')\phi^B(t')\tau(t',t)\right]
\end{eqnarray}
\end{widetext}
where 
\begin{equation}
\phi^{A,B}(t)=V^\dagger(\mathbf{x},t)K_U^{A,B}(\mathbf{x},\mathbf{y},t)V(\mathbf{y},t)
\end{equation}
is the characteristic kernel that reflects completely the structure and the properties of operator $\mathcal{O}^{A,B}(t)$. The last equation of Eq.~(\ref{eq:two-point}) is the generic
expression for mesonic two-point functions in the distillation method, where the perambulator $\tau(t,t')$ is universal and the information of the meson operators are completely encoded in $\phi^{A,B}(t)$. The disconnected part of the correlation function can be similarly derived in terms of $\phi^{A,B}(t)$ and $\tau(t,t')$ as follows
\begin{equation}
D_{AB}(t,t')=\mathrm{Tr}\left[\phi^A(t)\tau(t,t)\right]\mathrm{Tr}\left[\phi^B(t')\tau(t',t')\right].
\end{equation}
Finally, the full correlation function of $\mathcal{O}^A(t)$ and $\mathcal{O}^B(t')$ is 
\begin{equation}\label{eq:full-2pt}
G_{AB}(t,t')=C_{AB}(t,t')+D_{AB}(t,t').
\end{equation}
The equation (\ref{eq:full-2pt}) is the basic expression we apply to calculate the corresponding 
correlation functions of $1S$ and $1P$ charmonia.

There are subtleties in considering the contribution from the annihilation diagrams to the charmonium masses since there are two degenerate sea quark flavors in this work. This means there is a $SU(2)$ global flavor symmetry, named as the isospin symmetry in principle. Thus the connected part of the correlation functions are actually 
\begin{equation}
C_{ij}(t)=\frac{1}{2}C_{ij}^{ I=0}(t)+\frac{1}{2}C_{ij}^{I=1}(t)\equiv C_{ij}^{I=0}(t)
\end{equation}
if the flavor wave functions of $I=0,1$ are taken as $\frac{1}{\sqrt{2}}(\bar{c}c\pm\bar{c}'c')$. Similarly the disconnected part should be 
\begin{equation}
D_{ij}(t)=\frac{1}{2}D_{ij}^{I=0}(t),
\end{equation}
since gluons are isospin singlets and couple only with $I=0$ states. Therefore, the full correlation function of the $I=0$ states is 
\begin{equation}
G_{ij}^{I=0}(t)=C_{ij}(t)+2D_{ij}(t).
\end{equation}

A special attention should be paid to the $0^{++}$ channel. The quark loops in this channel have nonzero vacuum expectation values and should be subtracted. For this we adopt a `time-shift' subtraction scheme~\cite{Sun:2017ipk} by redefining the correlation function as 
\begin{equation}
\tilde{G}^{0^{++}}(t)=G^{0^{++}}(t)-G^{0^{++}}(t+\delta t)
\end{equation}
where the vacuum constant term cancels. In practice, we choose $\delta t=5$. The price we pay for this is a little lose of the signal-to-noise ratio.

\begin{table}[t]
	\centering \caption{\label{tab:gamma}The gamma matrix combinations $\Gamma$ for the $J^{PC}$ channels corresponding to $1S$ and $1P$ states. }
	\begin{ruledtabular}
		\begin{tabular}{ccccccc}
			$J^{PC}$    &  $0^{-+}$   & $1^{--}$   & $0^{++}$ & $1^{++}$           & $1^{+-}$ & $2^{++}$ \\
			$\Gamma$    &  $\gamma_5$ & $\gamma_i$ &  $I$     & $\gamma_5\gamma_i$ & $\epsilon_{ijk}\sigma_{jk}$& $|\epsilon_{ijk}|\gamma_j\nabla_k$ \\
			$1S, 1P$    &  $\eta_c$   & $J/\psi$   &  $\chi_{c0}$& $\chi_{c1}$ & $h_c$ &$\chi_{c2}$ \\
		\end{tabular}
	\end{ruledtabular}
\end{table}
\begin{figure*}[t!]
	\includegraphics[height=3.5cm]{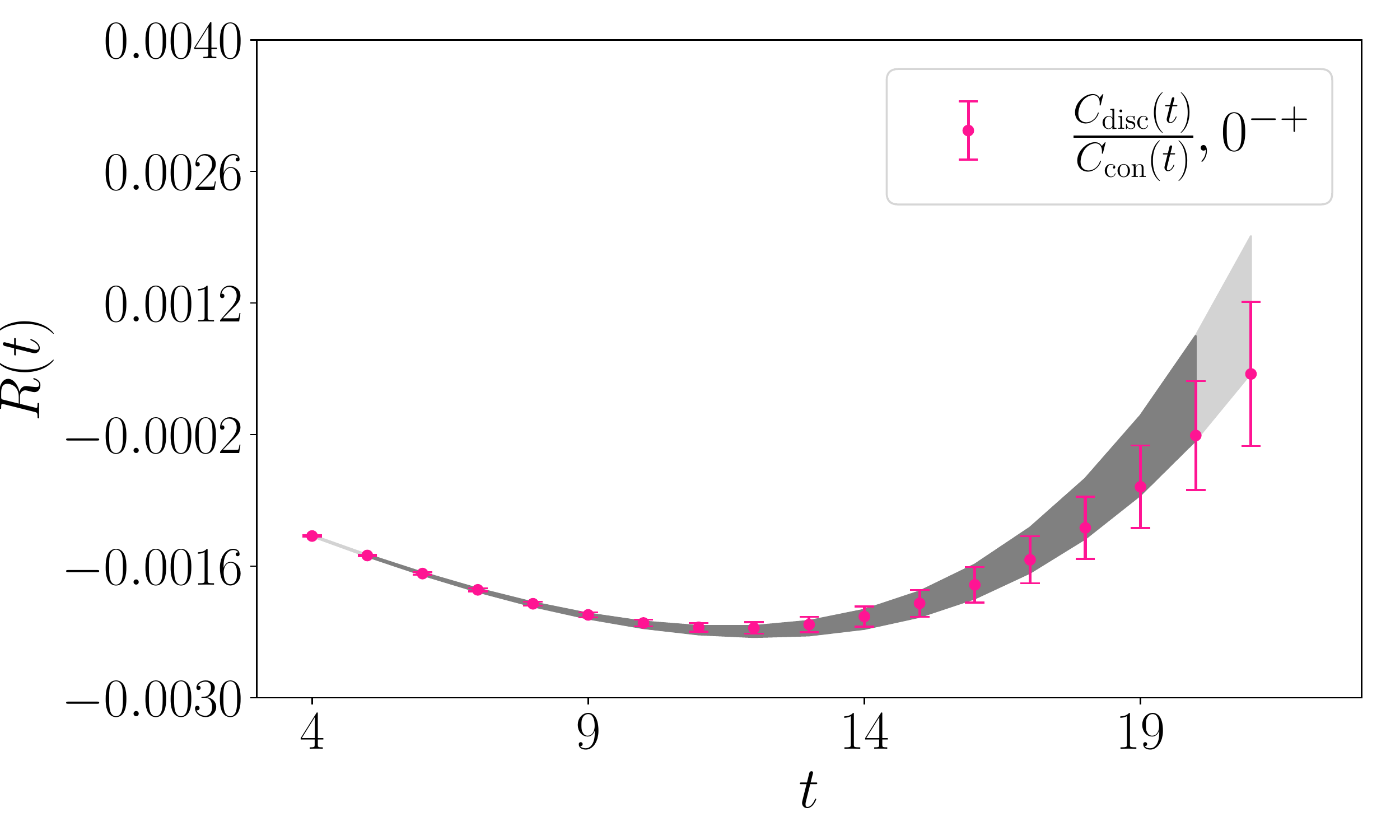}
	\includegraphics[height=3.5cm]{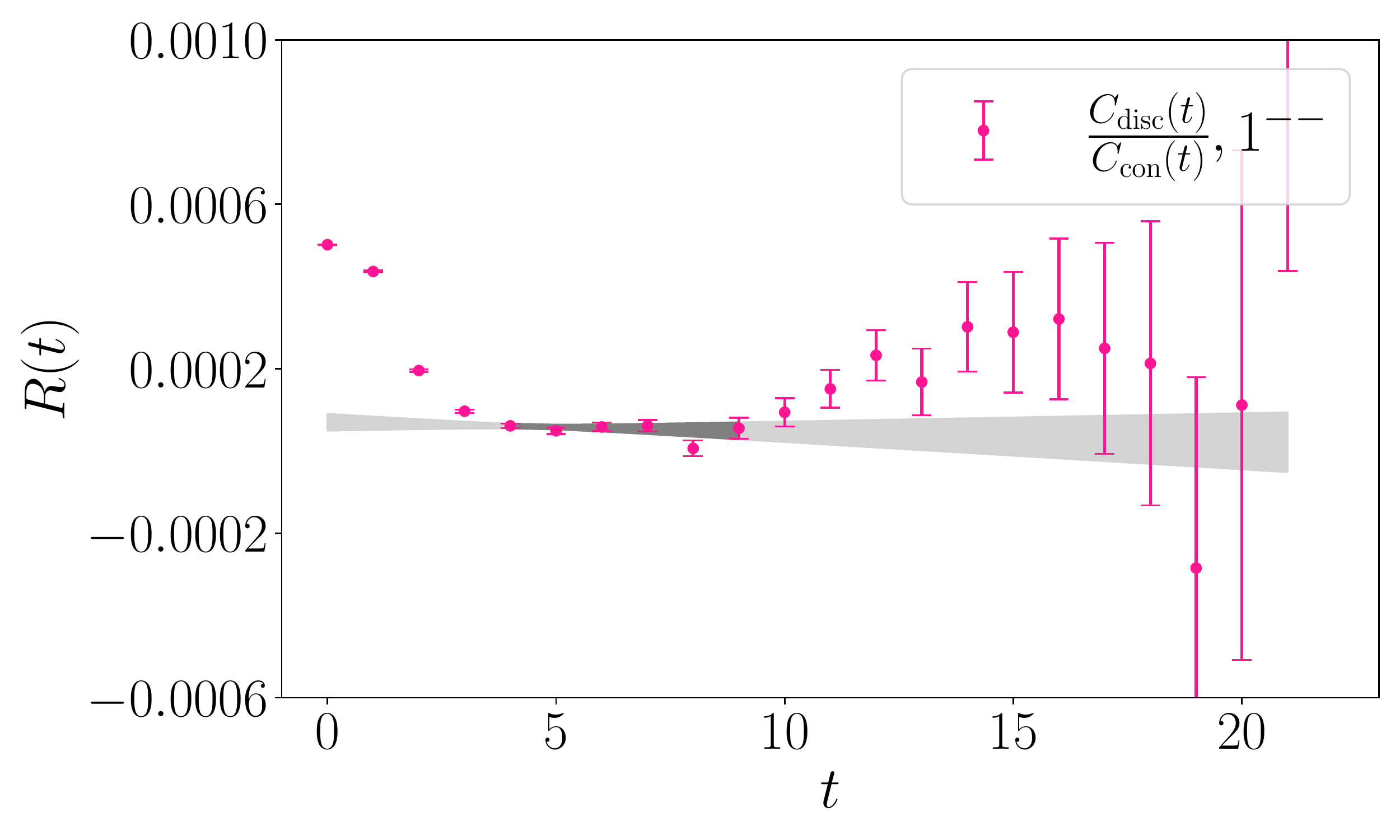}
	\includegraphics[height=3.5cm]{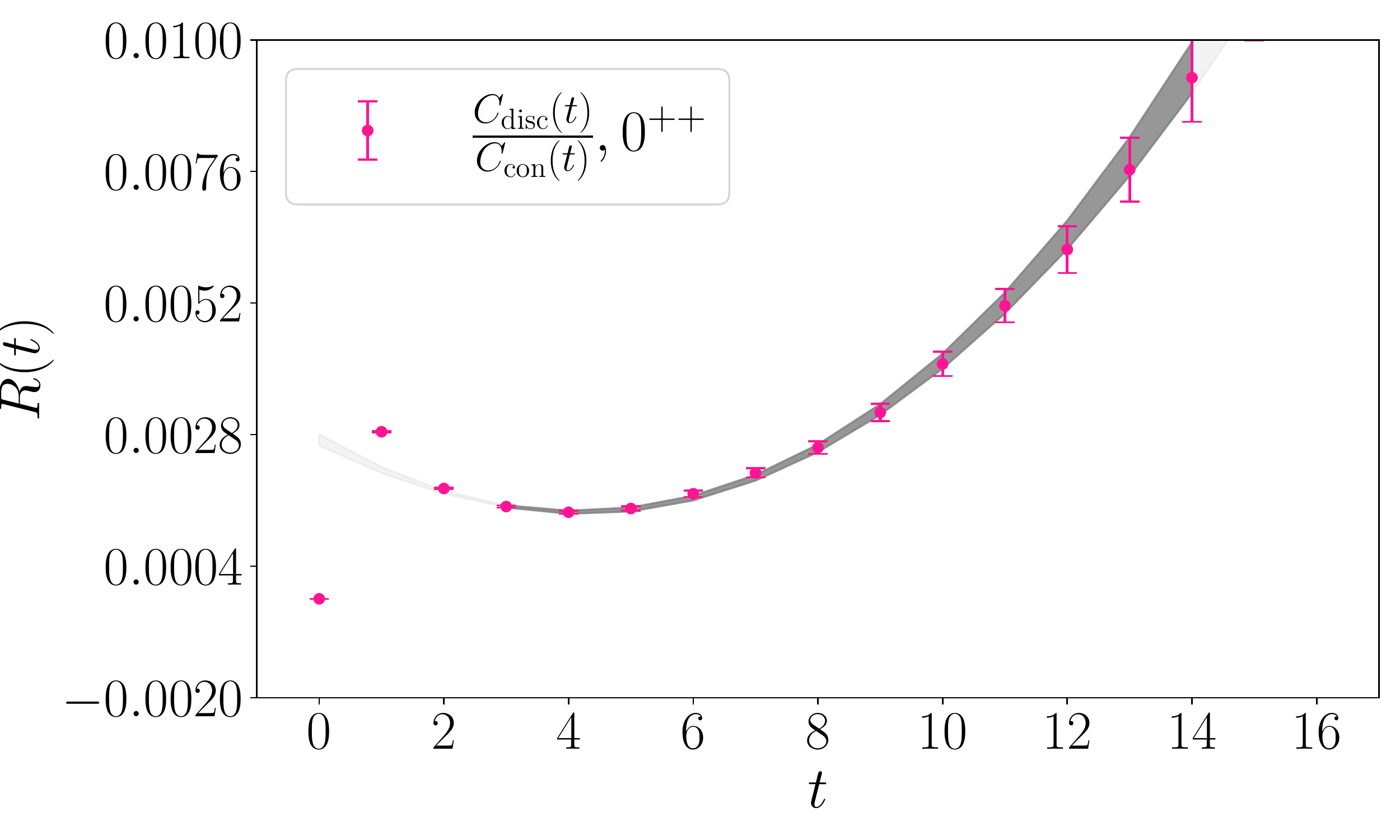}
	\includegraphics[height=3.5cm]{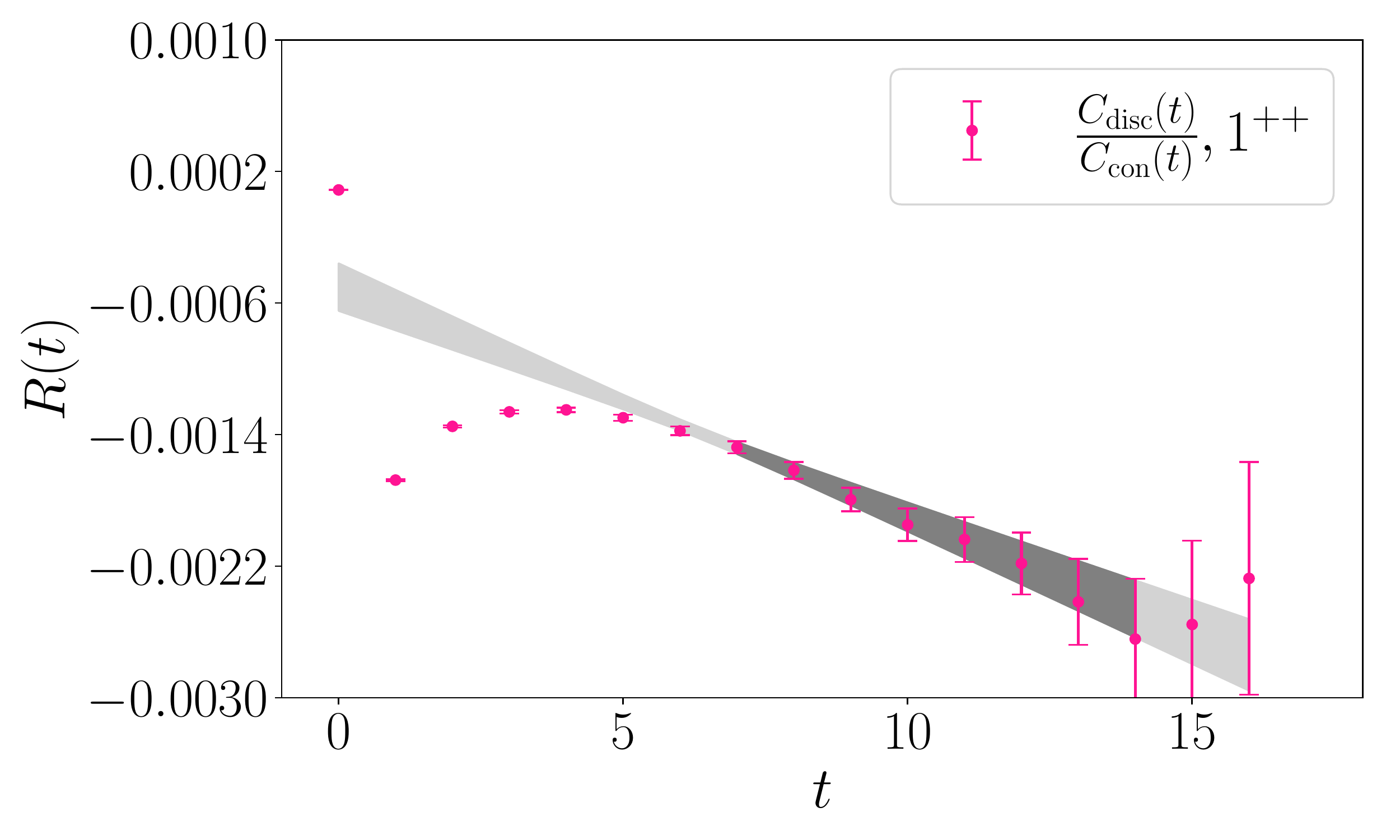}
	\includegraphics[height=3.5cm]{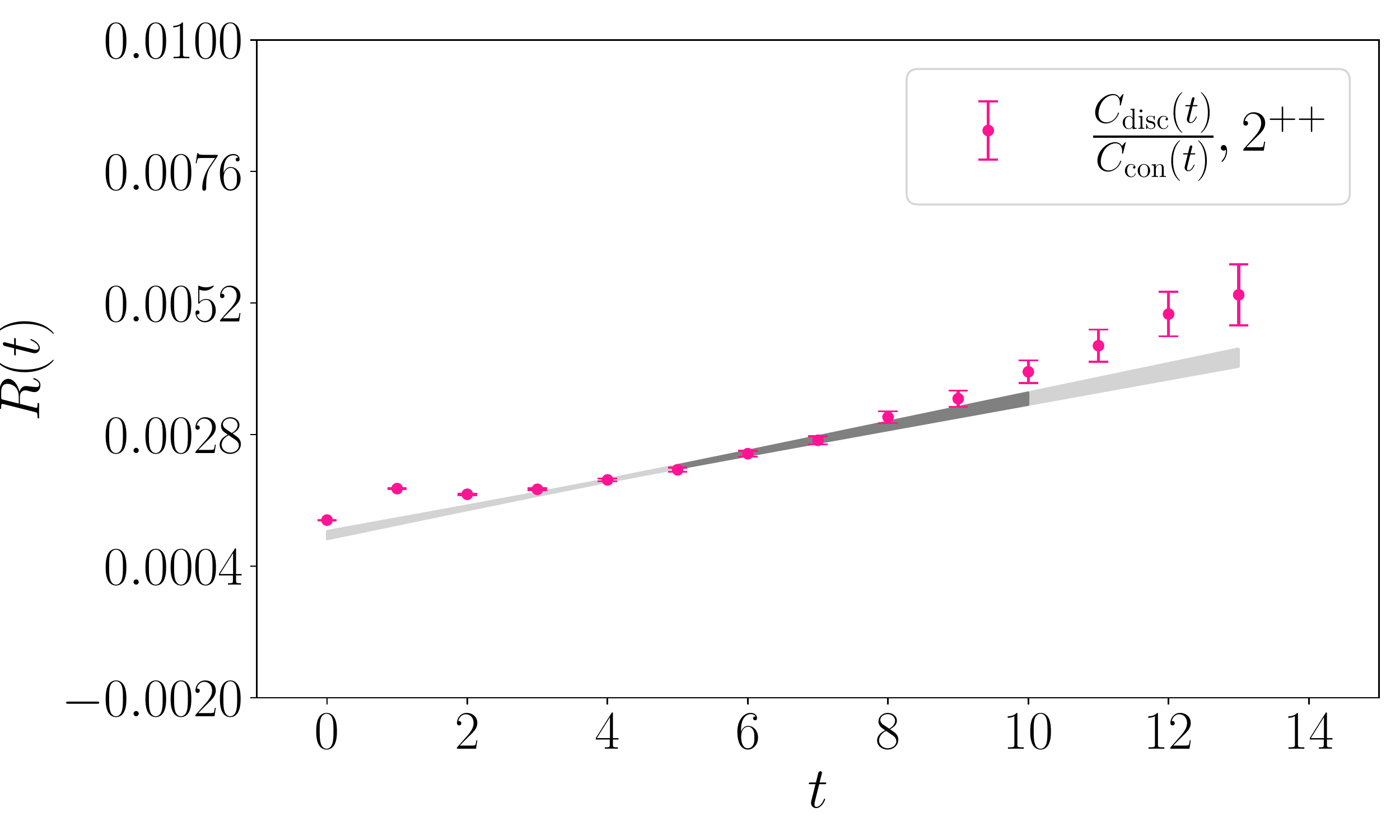}
	\includegraphics[height=3.5cm]{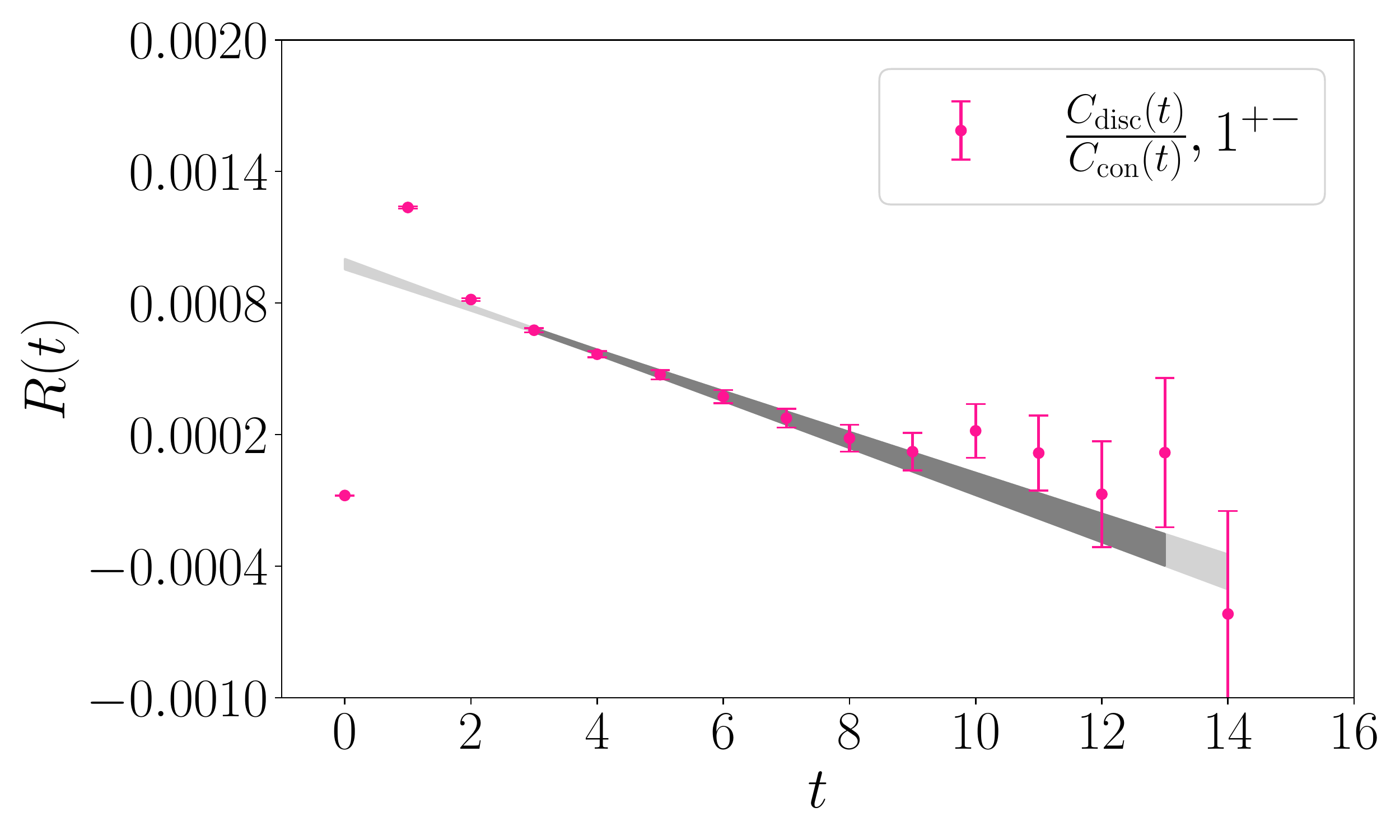}
	\caption{\label{fig:ratioE1} The ratio functions $R(t)$ for the six channels $J^{PC}=0^{-+},1^{--},(0,1,2)^{++}$ and $1^{+-}$ on gauge ensemble E1. The shaded curves with error bands are the fit results using the linear part of Eq.~(\ref{eq:linear-ratio}) for $1^{--},1^{+\pm}, 2^{++}$ channels and the results for $0^{\pm +}$ channels using Eq.~(\ref{eq:linear-ratio-ps}). The darker regions indicate the fit time range.}
\end{figure*}

\section{Annihilation diagram contribution in different channels}\label{sec:disc}
\begin{figure*}[t!]
	\includegraphics[height=3.5cm]{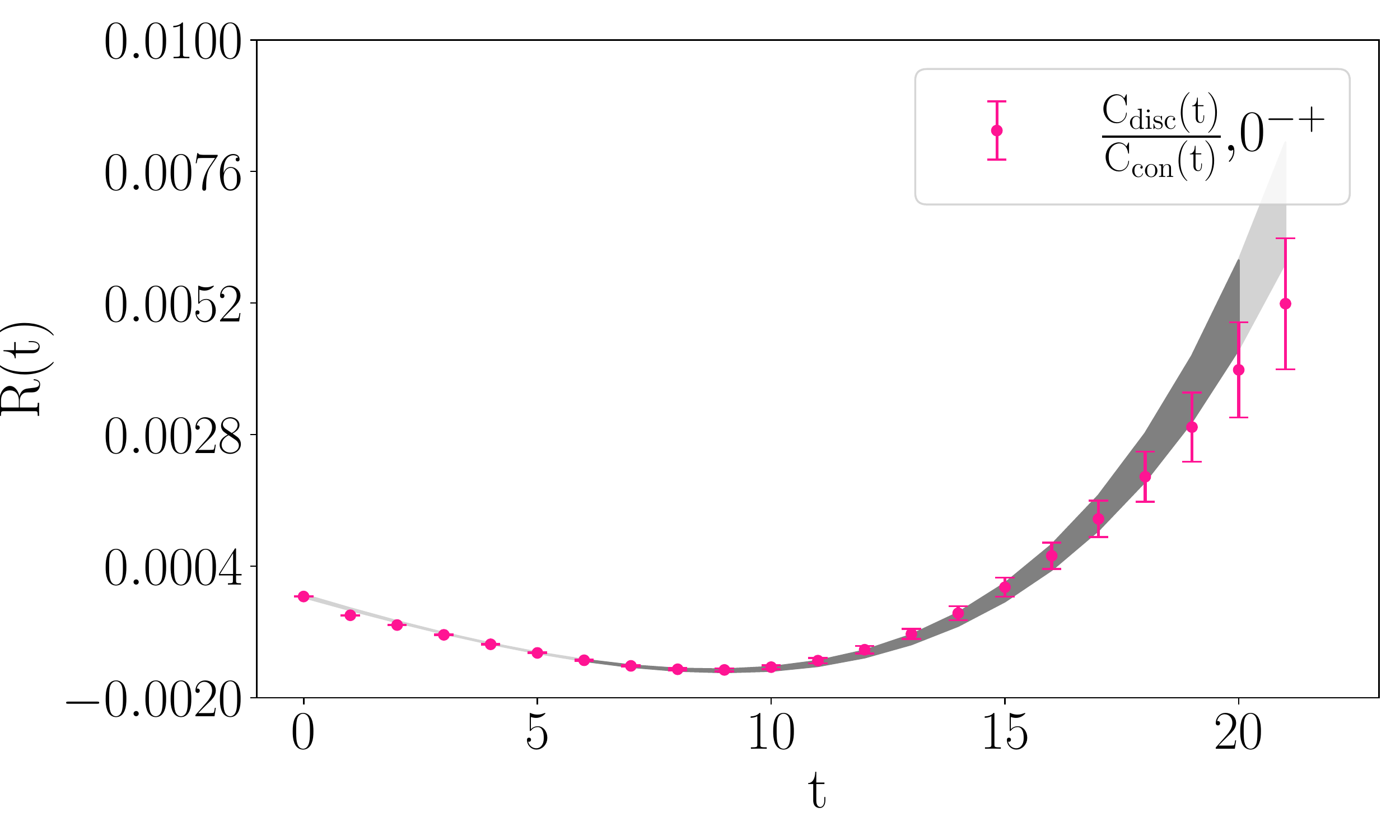}
	\includegraphics[height=3.5cm]{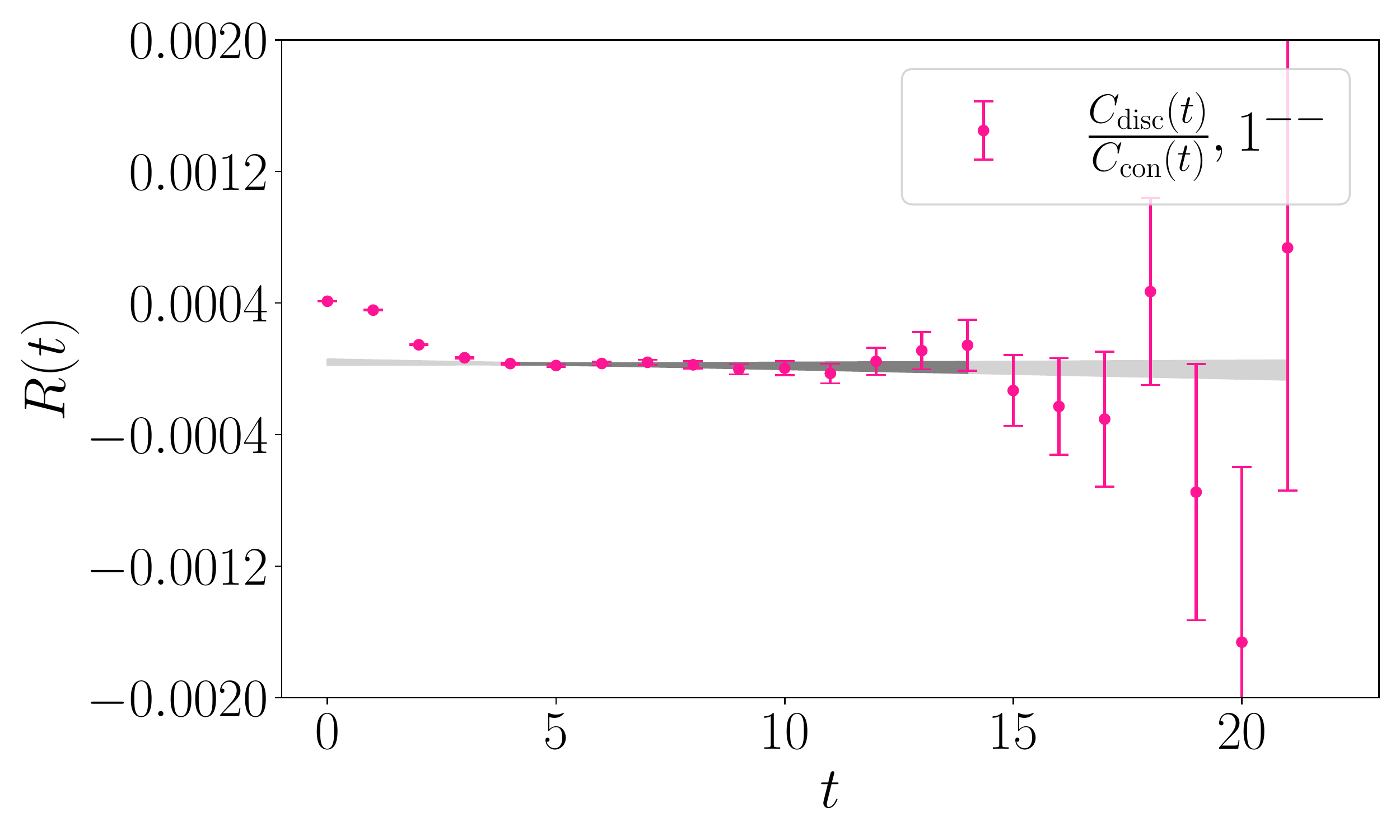}
	\includegraphics[height=3.5cm]{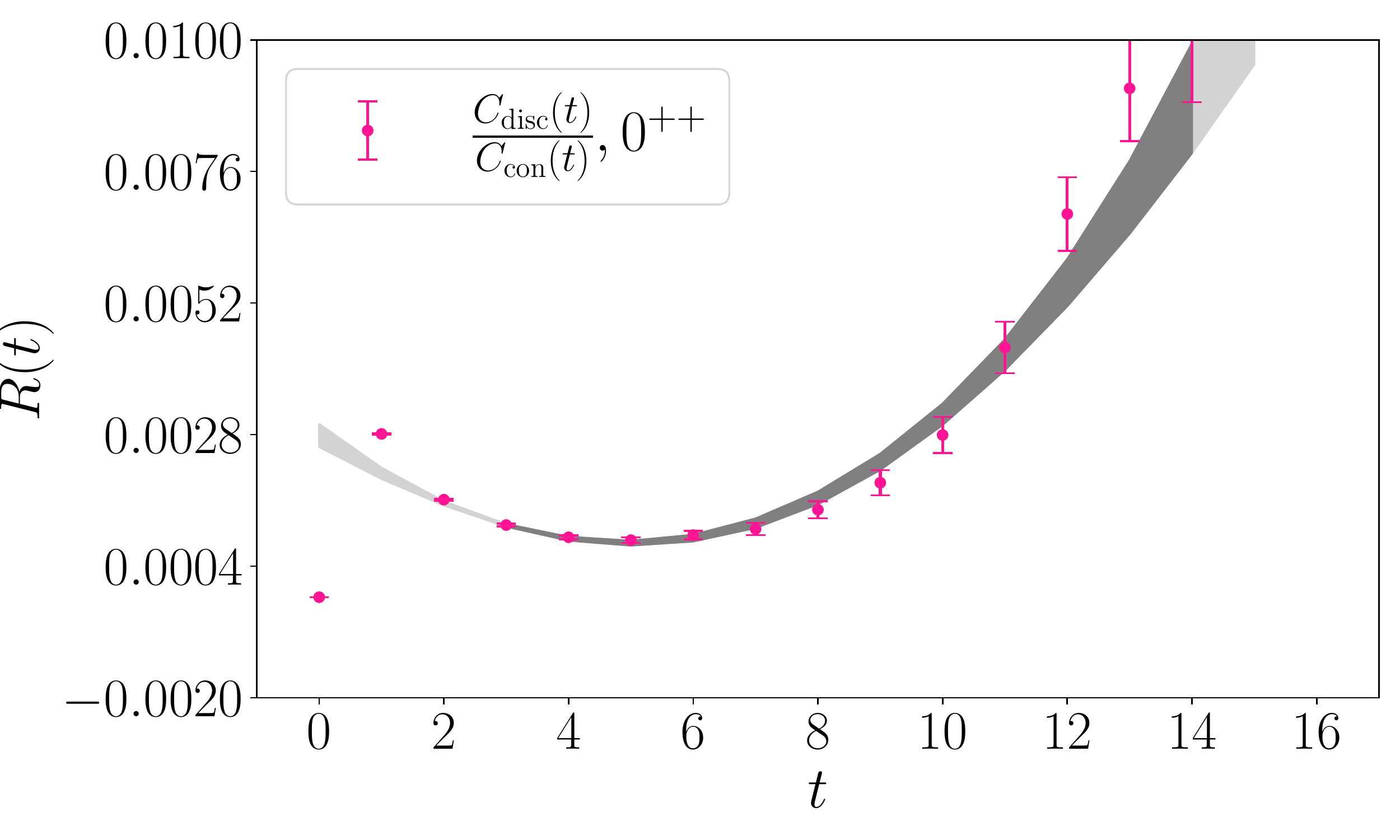}
	\includegraphics[height=3.5cm]{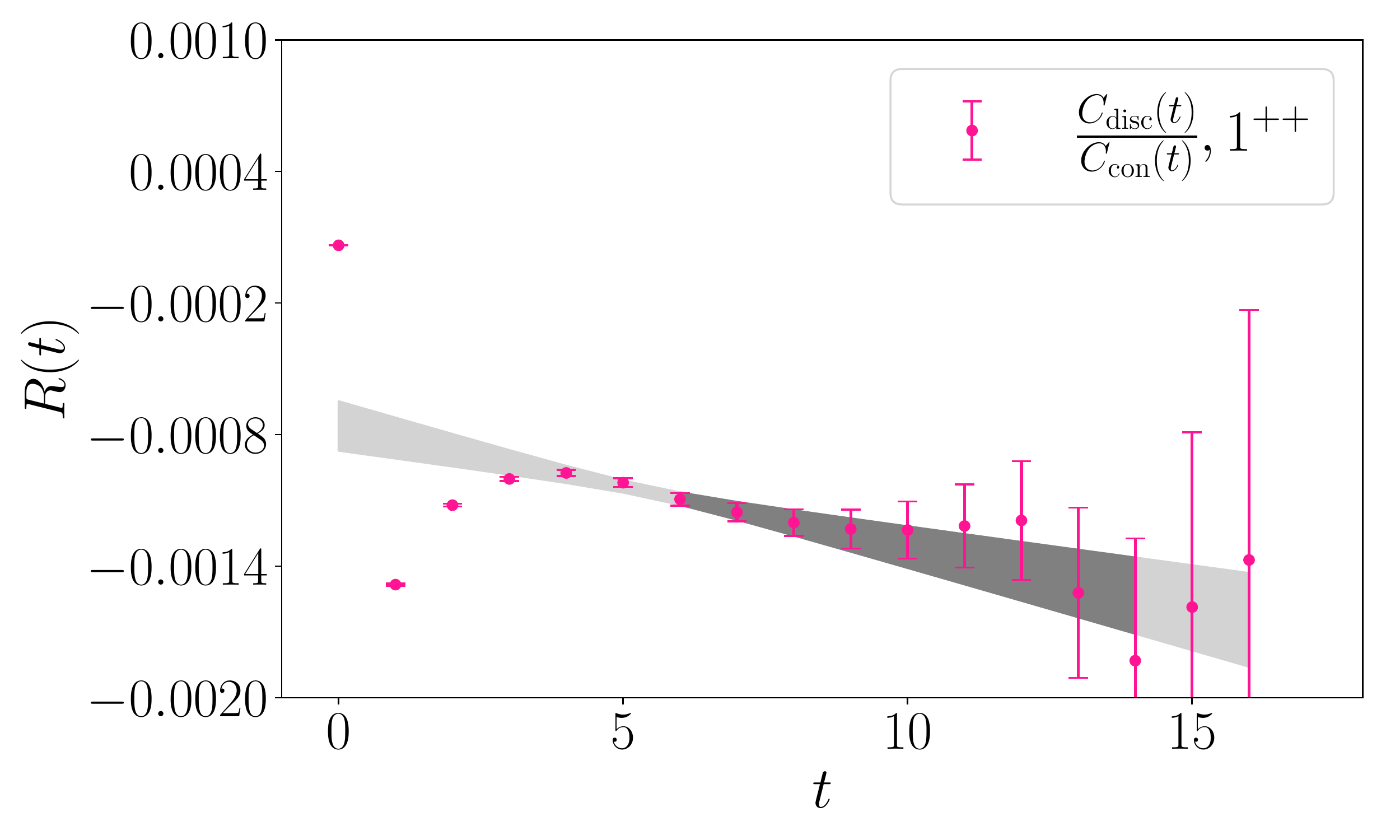}
	\includegraphics[height=3.5cm]{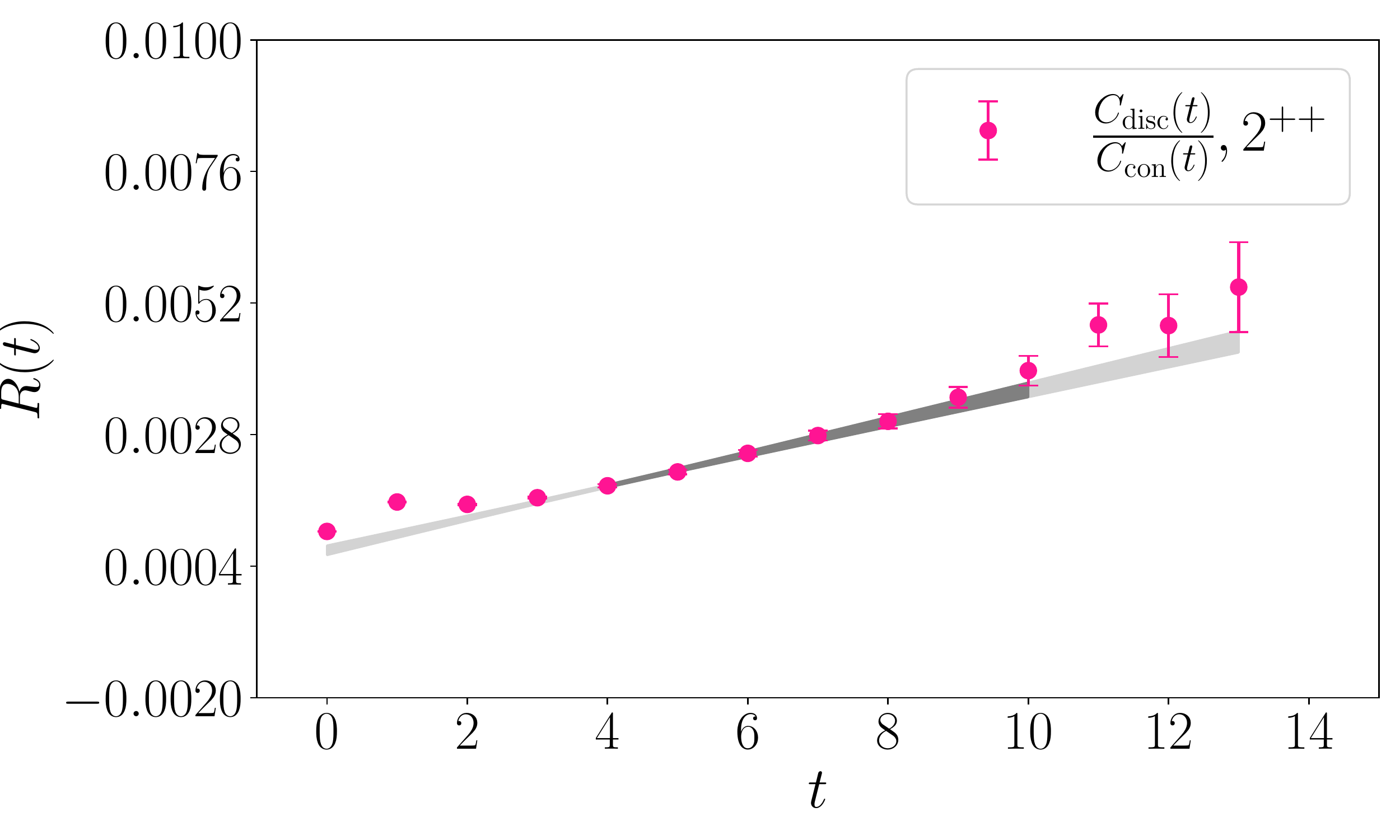}
	\includegraphics[height=3.5cm]{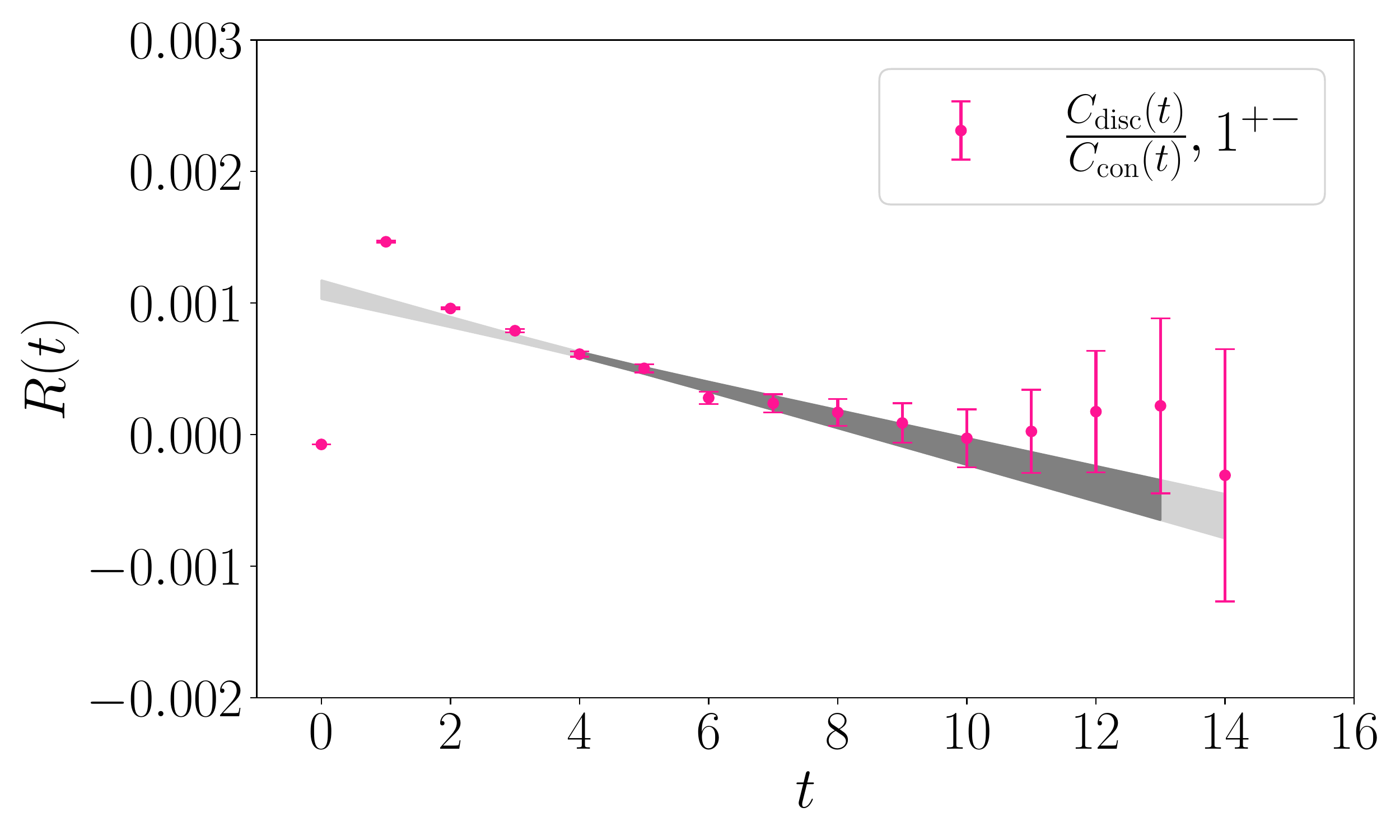}
	\caption{\label{fig:ratioE2} The ratio functions $R(t)$ for the six channels $J^{PC}=0^{-+},1^{--},(0,1,2)^{++}$ and $1^{+-}$ on gauge ensemble E2. The shaded curves with error bands are the fit results using the linear part of Eq.~(\ref{eq:linear-ratio}) for $1^{--},1^{+\pm}, 2^{++}$ channels and the results for $0^{\pm +}$ channels using Eq.~(\ref{eq:linear-ratio-ps}). The darker regions indicate the fit time range.}
\end{figure*}

As the first step, we compare the relative magnitudes of the contribution of the annihilation diagram to the connected counterpart in each channel, which 
is measured by the ratio for a specific channel labeled by $\Gamma$
\begin{equation}
R^\Gamma(t)\equiv \frac{2D_{11}^\Gamma(t)}{C_{11}^\Gamma(t)}
\end{equation}
where the subscript `11' refers to the correlation function of the $|r|/a_s=0$ operator  in the operator set of the $\Gamma$ channel. The ratios $R^\Gamma(t)$ at the two quark masses
are shown in Fig.~\ref{fig:ratioE1} and Fig.~\ref{fig:ratioE2} for channels with $J^{PC}=0^{-+}, 1^{--}, (0,1,2)^{++}$ and $1^{+-}$. It is seen the magnitudes of $R^\Gamma (t)$ are of order $\mathcal{O}(10^{-3}-10^{-5})$: The $R^\Gamma (t)$'s of $0^{-+}$, $0^{++}$ and $2^{++}$ channels are much larger than those of the other three channels. This is understandable since the charm quark loops in $0^{-+},0^{++}$ and $2^{++}$ channels are mediated by two gluons to the lowest order of QCD, while the charm quark loops in other three channels are mediated by at least three-gluon intermediate states. On the other hand, the disconnected part of $0^{-+}$ can be also enhanced by the QCD $U_A(1)$ anomaly. Especially, the $R^\Gamma(t)$ of $1^{--}$ is two order of magnitude smaller and turns out to be approximately independent of $t$. 

The contribution of the disconnected diagrams to charmonium masses can be estimated as follows. For convenience, we drop the superscript `$\Gamma$' temporarily in the following discussion. Usually the connected part of the charmonium correlation functions and the full correlation functions can be parameterized as 
\begin{eqnarray}\label{eq:twopoint-para}
C(t)&=&\sum\limits_i \frac{W_i}{2m_i} e^{-m_i t}\nonumber\\
G(t)&=&\sum\limits_i \frac{\tilde{W}_i}{2\tilde{m}_i} e^{-\tilde{m}_i t}.
\end{eqnarray}
Since $G(t)=C(t)+2D(t)$, the ratio of the disconnect part to the connected part can be expressed as 
\begin{equation}
R(t)=\frac{2D(t)}{C(t)}\equiv \frac{\sum\limits_i \frac{\tilde{W}_i}{2\tilde{m}_i} e^{-\tilde{m}_i t}}{\sum\limits_i \frac{W_i}{2m_i} e^{-m_i t}}-1. 
\label{eq:Rt}
\end{equation}

Intuitively, each mass term in the correlation function $C(t)$ is given by the propagator in the momentum 
space, namely, $W_i/(p^2-m_i^2)$. When the disconnected diagrams are considered, the propagator acquires a self-energy correction $-\Sigma(p^2)$, which is independent of the states. Thus the full propagator contributing to each term of $G(t)$ can be expressed as
\begin{widetext} 
\begin{eqnarray}\label{eq:full-prop}
\frac{\tilde{W}_i}{p^2+\tilde{m}_i^2}&\equiv&\frac{W_i}{p^2+m_i^2+\Sigma(p^2)}\nonumber\\
&=&\frac{W_i}{p^2+m_i^2}+\frac{\sqrt{W_i}}{p^2+m_i^2} (-\Sigma(p^2))\frac{\sqrt{W_i}}{p^2+m_i^2}
+ \frac{\sqrt{W_i}}{p^2+m_i^2} (-\Sigma(p^2))\frac{1}{p^2+m_i^2}(-\Sigma(p^2))\frac{\sqrt{W_i}}{p^2+m_i^2}+\cdots
\end{eqnarray}
\end{widetext}
with the relations
\begin{eqnarray}
\tilde{W}_i&=& W_i\left(1-\frac{d\Sigma(p^2)}{dp^2}|_{p^2=-\tilde{m}_i^2}\right)\equiv W_i(1+\epsilon_i)\nonumber\\
\tilde{m}_i^2 &=&m_i^2 +\Sigma(-\tilde{m}_i^2).
\end{eqnarray}
Based on the OZI rule and from the observation in Fig.~\ref{fig:ratioE1} and Fig.~\ref{fig:ratioE2}, it is safe to assume  $\epsilon_i\ll 1$ and $\delta m_i=\tilde{m}_i-m_i\ll m_i$. Thus we have 
\begin{equation}
\delta m_i = \frac{\Sigma(-\tilde{m}_i^2)}{2m_i}\approx \frac{m_1}{m_i}\left[\delta m_1+\Delta_{i}\left(1+\frac{\Delta_i}{2m_1}\right)\epsilon_1\right],
\end{equation}
where $\Delta_i=m_i-m_1$ has been defined and $\Sigma(p^2)$ is assumed to be varying very slowly with respect to $p^2$. The latter is obviously justified by the observation that $R(t)$ is usually of order $\mathcal{O}(10^{-3})$ or even smaller. If we assume $\epsilon_i$ is insensitive to $p^2$ in the energy range of interest, we can take the approximation $\epsilon_i\approx \bar{\epsilon}$ and $\delta m_i\approx \frac{m_1}{m_i}\delta m_1$, which imply
\begin{equation}
\tilde{\Delta}_i=\tilde{m_i}-\tilde{m}_1\approx \Delta_i\left(1-\frac{\delta m_1}{m_i}\right)\approx \Delta_i.
\end{equation}
Thus we have 
\begin{eqnarray}\label{eq:linear-ratio}
R(t)&\approx& -\delta m_1 t +\epsilon_1-\frac{\delta m_1}{m_1}\nonumber\\
&+&F(t)\frac{\delta m_1}{m_1} \sum\limits_{i>1} \frac{W_i m_1\Delta_i}{W_1 m_i^2}e^{-\Delta_i t}+\cdots
\end{eqnarray}
where 
\begin{equation}
F(t)=\left[1+\sum\limits_{i>1} \frac{W_i m_1}{W_1 m_i}e^{-\Delta_i t}\right]^{-1}.
\end{equation}
The last term in Eq.~(\ref{eq:linear-ratio}) will die out when $t$ increases. Therefore, if $R(t)$ shows up a linear dependence on $t$ in a time region, the contribution of the disconnected diagrams to the ground state mass, namely, $\delta m_1$, can be extracted by the slope of $R(t)$ in this region. 

From Fig.~\ref{fig:ratioE1} and Fig.~\ref{fig:ratioE2}, one can see that, for the $(1,2)^{++}$, $1^{--}$ and $1^{+-}$ channels, $R(t)$ do show up linear behaviors in different time regions. Therefore, in these time windows, we can drop the last term in Eq.~(\ref{eq:linear-ratio}) and fit the $R(t)$ using linear functions.  The fit results are shown in these plots by curves with error bands, where the bands in darker colors illustrate the fit time window $[t_\mathrm{min},t_\mathrm{max}]$ (the values of $t$ are in the lattice unit $a_t$ in the context). The fitted slopes $\delta m_1^\Gamma a_t$ and the $\chi^2/\mathrm{dof}$'s on the two ensembles (labeled as E1 and E2) are listed in Table~\ref{tab:disc-mass} along with the $\chi^2/\mathrm{dof}$ in the fitting time window $[t_\mathrm{min},t_\mathrm{max}]$ given in the table. It should be noted that the large errors of the ratio function $R(t)$ come from the disconnected diagram which becomes very noisy beyond $t\sim 10$. In the $1^{--}$ channel, $R(t)$ 
shows up good linear behavior in the time interval $t\in [4,9]$ (very close to a constant) on both ensemble E1 and E2, so we attribute the behavior of $R(t)$ in the 
interval $[9,14]$ to be mostly the statistical fluctuations. Actually, the data points in this interval deviate from the linear behaviors determined from the interval $t\in [4,9]$ by less than $2\sigma$. If we include the data points in the interval $t\in [10,14]$ to the fit, the central value does not change much by with a mildly larger $\chi^2/\mathrm{dof}$. Actually on E2 ensemble, all the data points in the interval $[4,14]$ converge to a single line. 

Obviously, the $\delta m_1$ of the $J/\psi$ is tiny and consistent with zero within the errors. This can be understood as the result of the strong suppression based on the Okubo-Zweig-Iizuka rule (OZI rule)~\cite{Okubo:1963fa,Mandula:1970wz,tHooft:1976rip}, since the two charm quark loops are mediated by at least three intermediate gluons.  For $\chi_{c1}$ and $h_c$, the mass shifts due to the charm annihilation effect are small but non-zero. This is understandable since the contribution of two-gluon intermediate states is suppressed to some extent due to the Landau-Yang theorem~\cite{Landau:1948kw,Yang:1950rg}. It is surprising that this kind of mass shift of $\chi_{c2}$ is relatively large and negative. The reason for this is not clear yet. Intuitively, lattice QCD studies predict that the lowest tensor glueball mass is approximately 2.2-2.4 GeV~\cite{Morningstar:1999rf,Chen:2005mg,Sun:2017ipk,Richards:2010ck,Gregory:2012hu}, which is lower than the $\chi_{c2}$ mass and should result in positive mass shift if $\chi_{c2}$ and the tensor glueball can mix.  

\begin{table*}[t]
	\centering \caption{\label{tab:disc-mass}The mass shifts $a_t\delta m_1$ for all the six channels. The fit time ranges $[t_\mathrm{min},t_\mathrm{max}]$ and the values of the $\chi^2$ per degree of freedom ($\chi^2/\mathrm{dof}$) are also presented.}
	\begin{ruledtabular}
		\begin{tabular}{ccccccc}
			&  $\eta_c(0^{-+})$   & $J/\psi(1^{--})$   & $\chi_{c0}(0^{++})$ & $\chi_{c1}(1^{++})$   & $\chi_{c2}(2^{++})$    & $h_c(1^{+-})$ \\
			$[t_\mathrm{min},t_\mathrm{max}]$ & $[5,20]$ & $[4,9]$ & $[3,14]$ & $[7,14]$ & $[5,10]$ & $[3,13]$ \\
			$\delta m_1^\Gamma a_t$ (E1)    &  0.00031(1) &   $\sim10^{-6}$    & 0.007(5) &  0.00014(2) & -0.00031(3) & 0.00010(1)\\
			$\chi^2/\mathrm{dof}$ &   1.5  & 0.78 &  0.62 & 0.59 & 0.82 & 0.71\\
			$[t_\mathrm{min},t_\mathrm{max}]$ & $[6,20]$ & $[4,14]$ & $[3,14]$ & $[6,14]$ & $[4,10]$ & $[4,13]$ \\
			$\delta m_1^\Gamma a_t$ (E2)    &  0.00032(2) &   $\sim10^{-6}$    & 0.003(4) &  0.00006(2) & -0.00029(2) & 0.00012(2)\\
			$\chi^2/\mathrm{dof}$ &   1.4  & 0.59 &  1.2 & 0.64 & 0.71 & 1.3\\
		\end{tabular}
	\end{ruledtabular}
\end{table*}

The situation of the scalar and the pseudoscalar channels is a little more complicated. It is seen in Fig.~\ref{fig:ratioE1} and Fig.~\ref{fig:ratioE2} that the ratio functions $R(t)$ in these two channels increase rapidly when $t$ is large. This observation signals that there may be lighter states than the ground state charmonium contributing to the correlation function $G(t)$ in the scalar and pseudoscalar channels. Actually, lattice QCD calculations predict that the masses of the lowest scalar and the pseudoscalar 
glueballs are 1.5-1.7 GeV and 2.4-2.6 GeV~\cite{Morningstar:1999rf,Chen:2005mg,Sun:2017ipk,Richards:2010ck,Gregory:2012hu}, respectively. When the disconnected diagrams are considered, these glueballs can contribute as intermediate states to the correlation function $G(t)$. Therefore, more mass terms should be added into the expression of $G(t)$ (Eq.~(\ref{eq:twopoint-para})) to account for the possible contribution from glueballs. Consequently, Eq.~(\ref{eq:Rt}) for these two channels may be modified to 
\begin{equation}\label{eq:linear-ratio-ps}
R'(t)\approx R(t)+W_g e^{\Delta_g t}
\end{equation}
where $\Delta_g=m_1-m_G>0$ is a parameter resembling the mass difference between the ground state glueball and the corresponding charmonium state. Of course, one-glueball or multi-glueball states (if they exist) can also appear in  other channels, but Fig.~\ref{fig:ratioE1} and Fig.~\ref{fig:ratioE2} show that their contributions have no 
sizable effects and can be ignored in practice. 

We tentatively use the functions form of Eq.~(\ref{eq:linear-ratio-ps}) to fit $R(t)$ of the scalar and the pseudoscalar channels, with $R(t)$ in this equation being approximated by a linear function. The fit results are illustrated in Fig.~\ref{fig:ratioE1} and Fig.~\ref{fig:ratioE2} by curves with error bands, where one can see that the function form describes the data very well in large 
time ranges. This manifests the efficacy and the necessity of the second term in Eq.~(\ref{eq:linear-ratio-ps}). The mass shift $\delta m_1$ of the ground state pseudoscalar charmonium $\eta_c$ is determined to be +3.0(1) MeV for E1 ensemble and +3.1(2) MeV for E2 ensemble, respectively. $\delta m_1$ is converted into the value in physical units using the temperal lattice spacing $a_t^{-1}=9.62$ GeV. This result is consistent with the previous result in Ref.~\cite{Levkova:2010ft} and the result $\delta m_1=3.9(9)$ MeV derived from the glueball-charmonium mixing mechanism~\cite{Zhang:2021xvl}.

The values of the $\chi_{c0}$ mass shift are fitted to be 67(48) MeV for 
E1 ensemble and 29(38) MeV for E2 ensemble, respectively. The results for $\chi_{c0}$ have large errors due to nonexistence or the shortness of the linear behavior. Due to the large error, we cannot get a reliable final result of the $\delta m_1$ for $\chi_{c0}$. 

 Using the lattice spacing $a_t^{-1}\approx 9.62$ GeV, we get the mass shifts $\delta m_1$ as follows:
\begin{eqnarray}\label{eq:deltam}
\mathrm{E1}:~~~&& \delta m_1 (0^{-+})=3.0(1)~~\mathrm{MeV}\nonumber\\
            && \delta m_1 (1^{++})=1.3(2)~~\mathrm{MeV}\nonumber\\
            && \delta m_1 (2^{++})=-3.0(3)~~\mathrm{MeV}\nonumber\\
            && \delta m_1 (1^{+-})=1.0(1)~~\mathrm{MeV}\nonumber\\
\mathrm{E2}:~~~&& \delta m_1 (0^{-+})=3.1(2)~~\mathrm{MeV}\nonumber\\
            && \delta m_1 (1^{++})=0.6(2)~~\mathrm{MeV}\nonumber\\
            && \delta m_1 (2^{++})=-2.8(2)~~\mathrm{MeV}\nonumber\\
            && \delta m_1 (1^{+-})=1.2(2)~~\mathrm{MeV}.
\end{eqnarray}
Since the fitted $\delta m_1(1^{--})$ is very small, we do not quote its value here but only say that the charm annihilation effects in $1^{--}$ are negligible. The mass shifts $\delta m_1$ from E1 and E2 ensembles are compatible with each other in most channels except for the $1^{++}$ channel, where the deviation of $\delta m_1$ values on the two ensembles is larger than $3\sigma$. The reason for this discrepancy is not clear yet and should be investigated in the future.     
 
\section{The center of gravity mass of $P$-wave charmonium}\label{sec:COG}
Now that we have calculated the perambulators of charm quarks on the two gauge ensembles of large statistics, we would like to take a look at the charmonium spectrum. In this work, we will focus on the lowest two states in $J^{PC}=0^{-+},1^{--},(0,1,2)^{++}$ and $1^{-+}$ channels. In the previous sections we have shown that the inclusion of the disconnected diagrams does not change much the masses of charmonium states, so in the following discussions we only consider the connected part of the correlation functions. For each $J^{PC}$ quantum number, we will first build an operator set, then calculate the correlation matrix of the operators in this set. After that, we will solve the generalized eigenvalue problem (GEVP) to obtain the optimized operators that couple most to specific states. What follows are the technique details.  
\vspace{0.5cm}
\subsection{Optimized operators through variational method}
As we addressed in Sec.~\ref{sec:distillation}, the distillation method automatically provides the gauge covariant smearing function $\Phi^{ab}(\mathbf{x},\mathbf{y})$ for quark fields on each timeslice. To be specific, the smeared quark field $\hat{c}(\mathbf{x},t)$ can be obtained by $\hat{c}^a(\mathbf{x},t)=\Phi^{ab}(\mathbf{x},\mathbf{y}) c^b(\mathbf{y},t)$, where the duplicated spatial coordinate $\mathbf{y}$ means the summation over the space volume. Thus the quark bilinear operators for meson states in this study can be built in terms of the smeared charm quark field $\hat{c}(x)$.
We introduce the spatially extended operators for each channel as 
\begin{equation}
\mathcal{O}(r,t)=\frac{1}{N_r}\sum\limits_{\mathbf{x},|\mathbf{r}|=r}\bar{\hat{c}}(\mathbf{x},t)K_U^\Gamma(\mathbf{x,x+r};t)\hat{c}(\mathbf{x+r},t)
\end{equation}
where $N_r$ is the multiplicity of $\mathbf{r}$ with $|\mathbf{r}|=r$, and 
\begin{equation}
K_U^\Gamma(\mathbf{x,x+r};t)=\Gamma \mathcal{P}e^{-ig\int_{\mathbf{x}}^{\mathbf{x+r}} \mathbf{A}\cdot d\mathbf{r}}\equiv \Gamma \mathcal{L}(\mathbf{x,x+r};t)
\end{equation}
with $\Gamma$ being the specific combination of $\gamma$-matrix that gives the right quantum number
of each of the $1S$ and $1P$ state (The explicit expressions of $\Gamma$'s are tabulated in Table~\ref{tab:gamma}), and $\mathcal{L}(\mathbf{x,x+r};t)$ being the gauge transportation operator from $(\mathbf{x},t)$ to $(\mathbf{x+r},t)$ . Obviously, $\mathcal{O}(r,t)$ is gauge invariant. Thus one can get the kernel function 
\begin{equation}
	\phi^{r}(t)=\frac{1}{N_r}\sum\limits_{\mathbf{x},|\mathbf{r}|=r}V^\dagger(\mathbf{x},t)K_U^\Gamma(\mathbf{x,x+r};t)V(\mathbf{x+r},t).
\end{equation}
Because the disconnected part is far more smaller than connected part and has little effect on present discussion, the correlation function $G_{ij}(t)$ can be approixmated by 
their connected part as 
\begin{eqnarray}
G_{ij}(t)&=&\frac{1}{T}\sum\limits_{t_0=0}^{T-1} \langle \mathcal{O}(r_i,t+t_0)\mathcal{O}(r_j,t_0)\rangle\nonumber\\
&\approx& C_{ij}(t)
\end{eqnarray}
where $C_{ij}(t)$ is the connected part
\begin{eqnarray}
C_{ij}(t)&=&-\frac{1}{T}\sum\limits_{t_0=0}^{T-1}\left\langle\mathrm{Tr}\left[\phi^{r_j}(t+t_0)\tau(t+t_0,t_0)\right.\right.\nonumber\\
&&\times \left.\left.\phi^{r_i}(t_0)\tau(t_0,t+t_0)\right]\right\rangle.
\end{eqnarray}
\begin{figure}
 	\includegraphics[height=5cm]{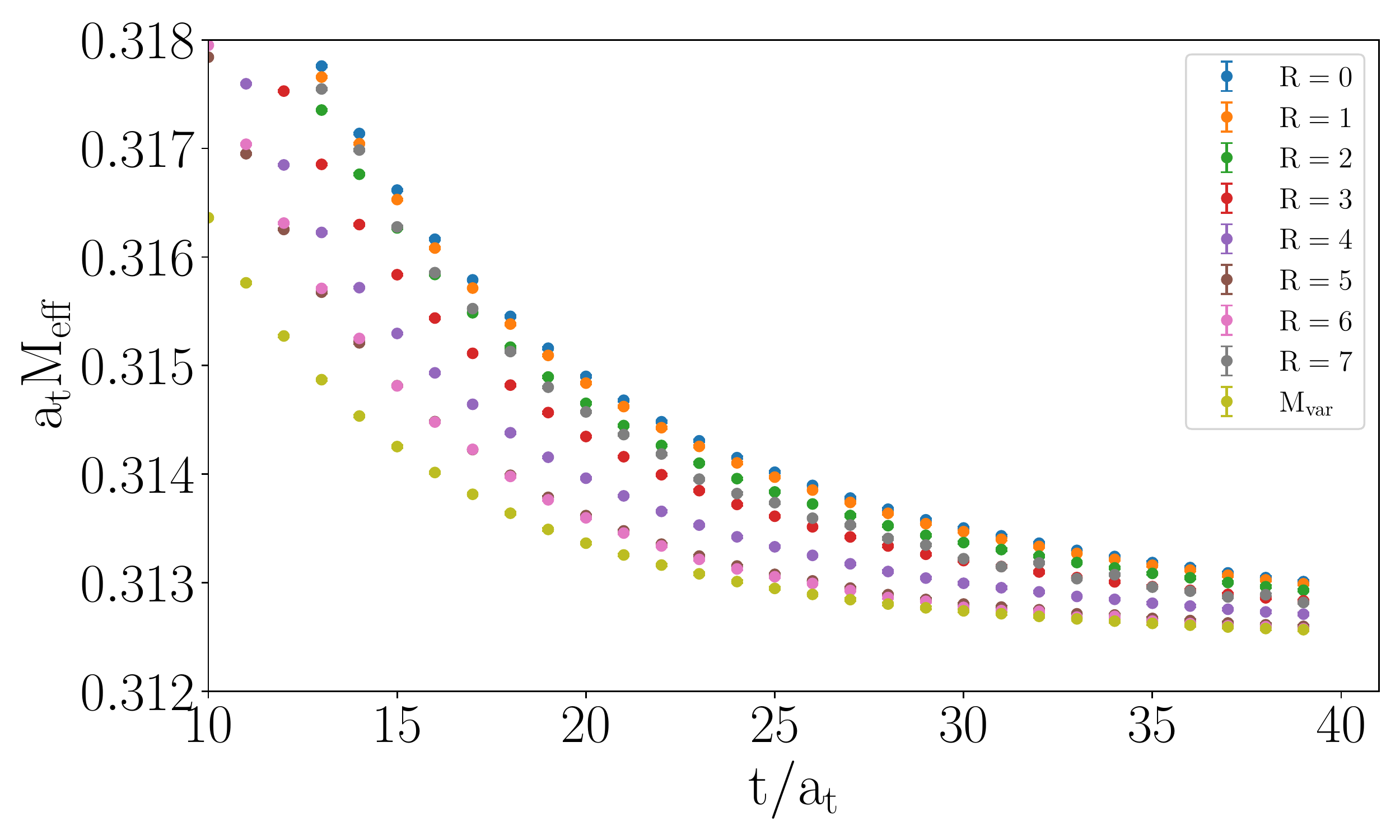}
	\caption{\label{fig:VarR} Effective mass of these correlators with different $r$ (labelled by $R=r/a_s$). $\mathrm{M_{var}}$ labels the effective mass of ground state optimized operator obtained from $3\times 3$ variational analysis.}
\end{figure}

\begin{figure*}[t!]
	\includegraphics[height=3.5cm]{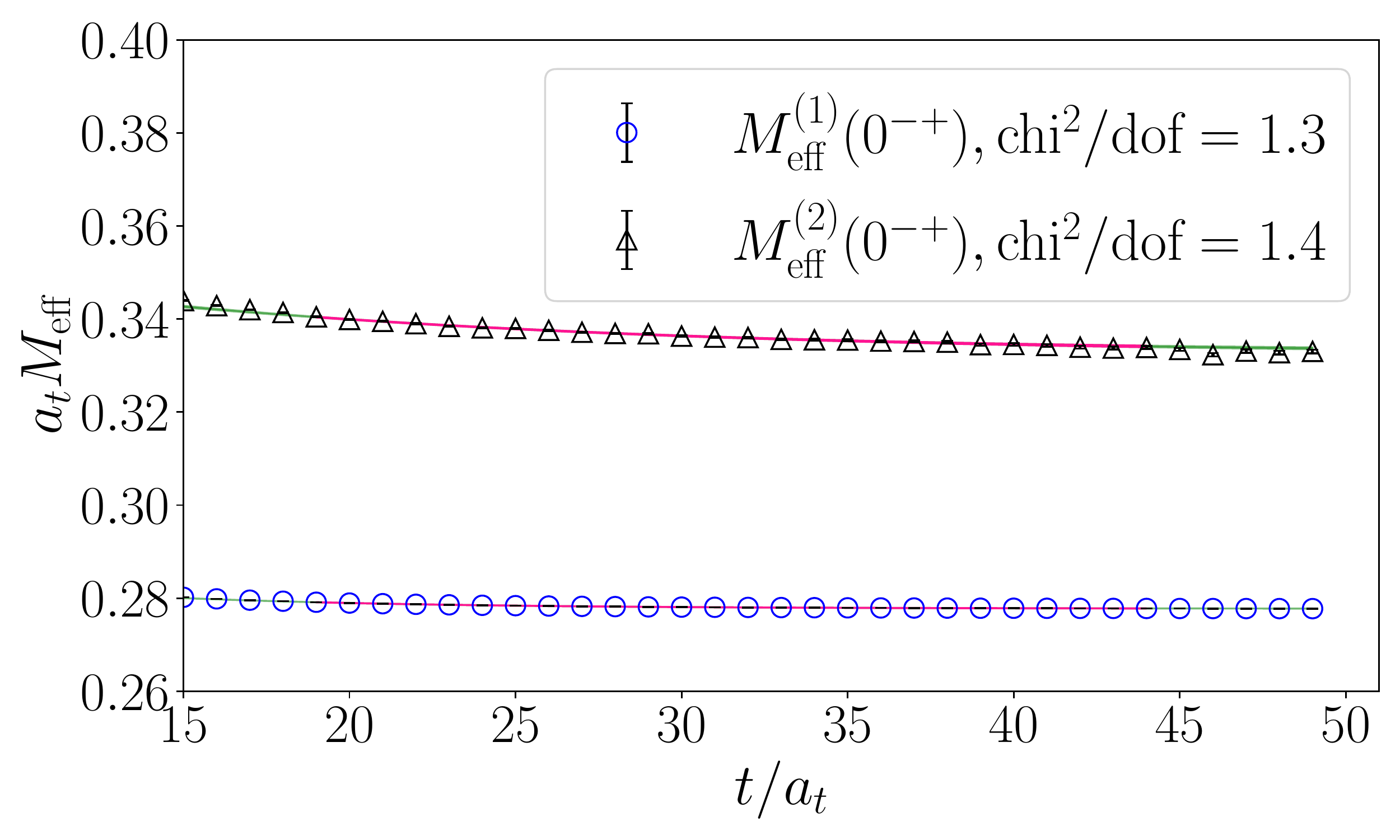}
	\includegraphics[height=3.5cm]{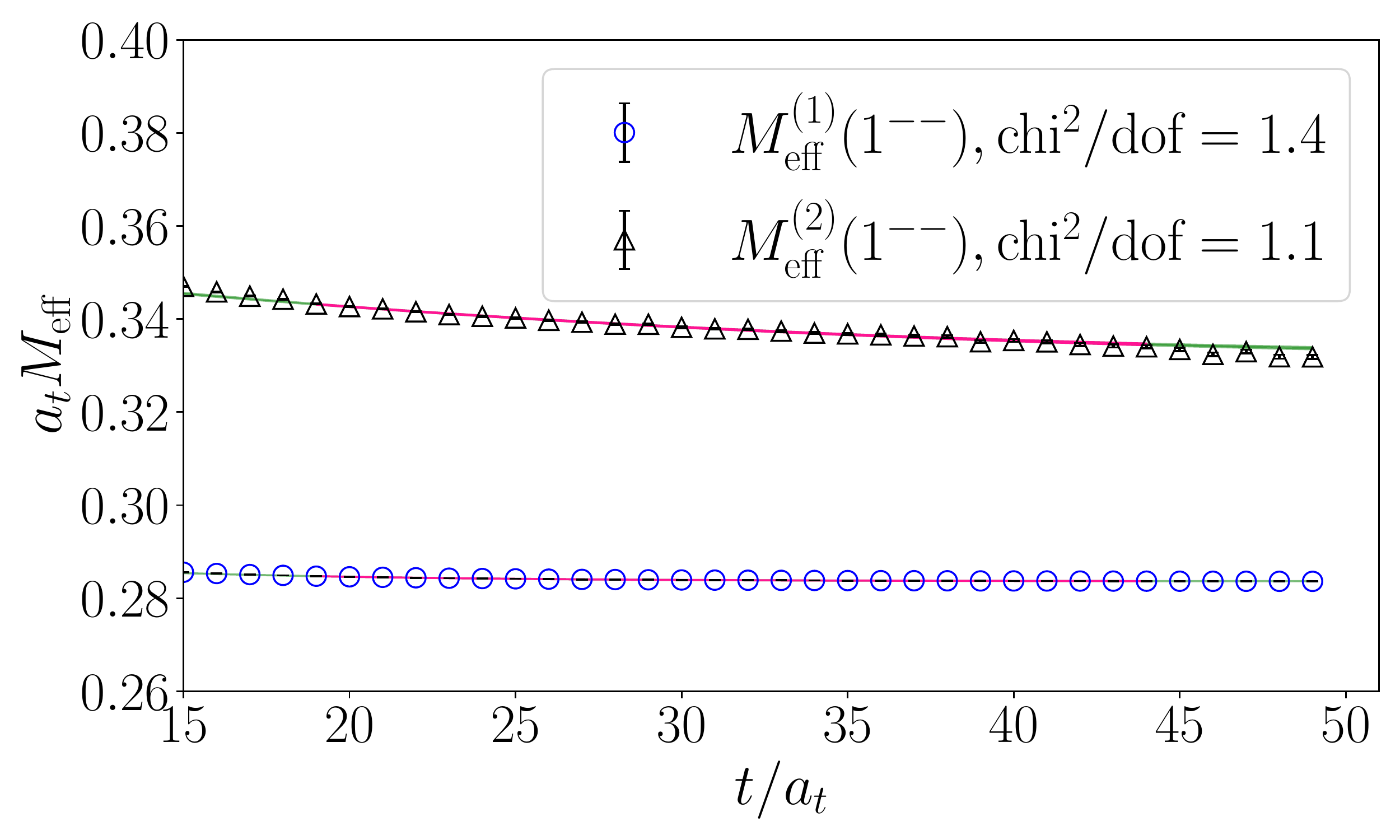}
	\includegraphics[height=3.5cm]{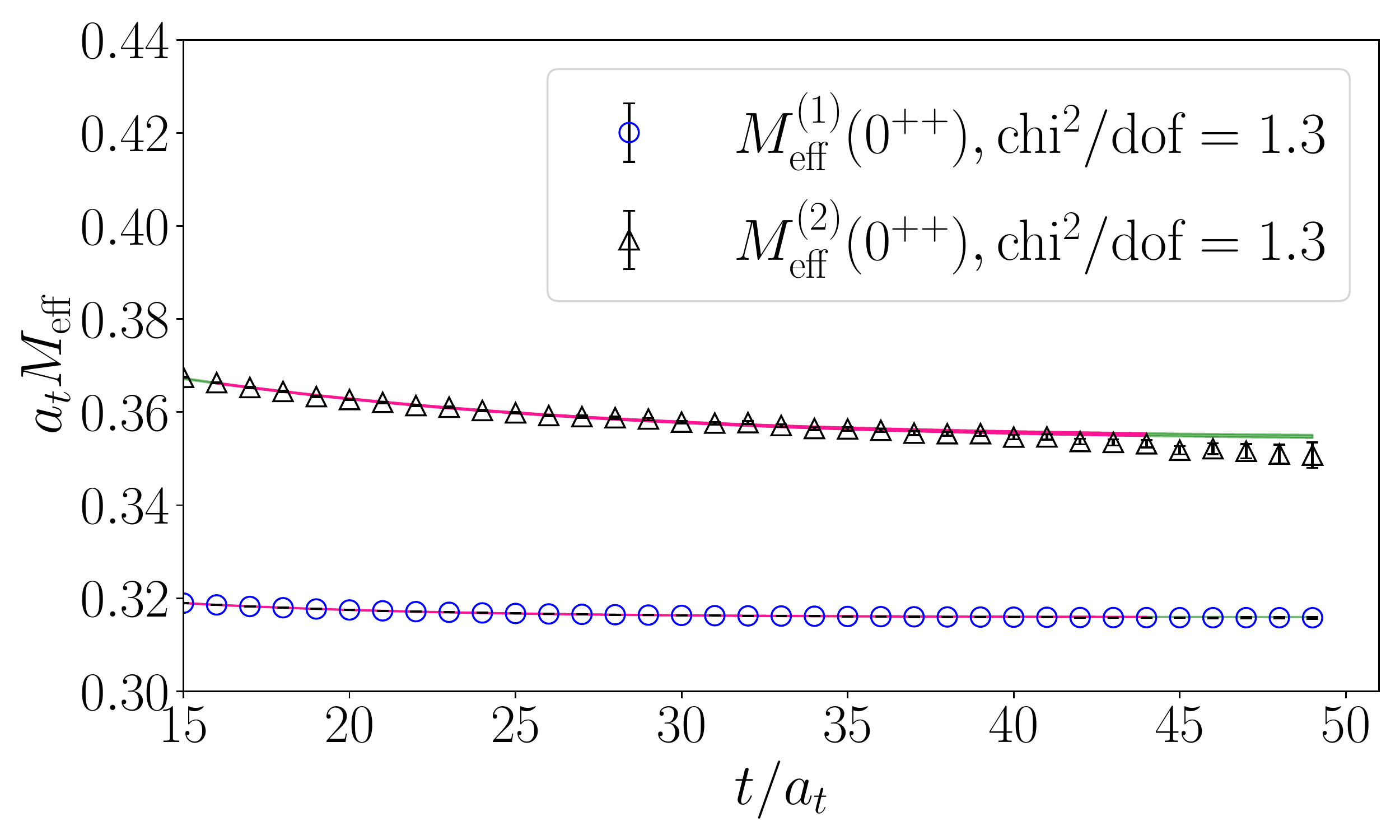}
	\includegraphics[height=3.5cm]{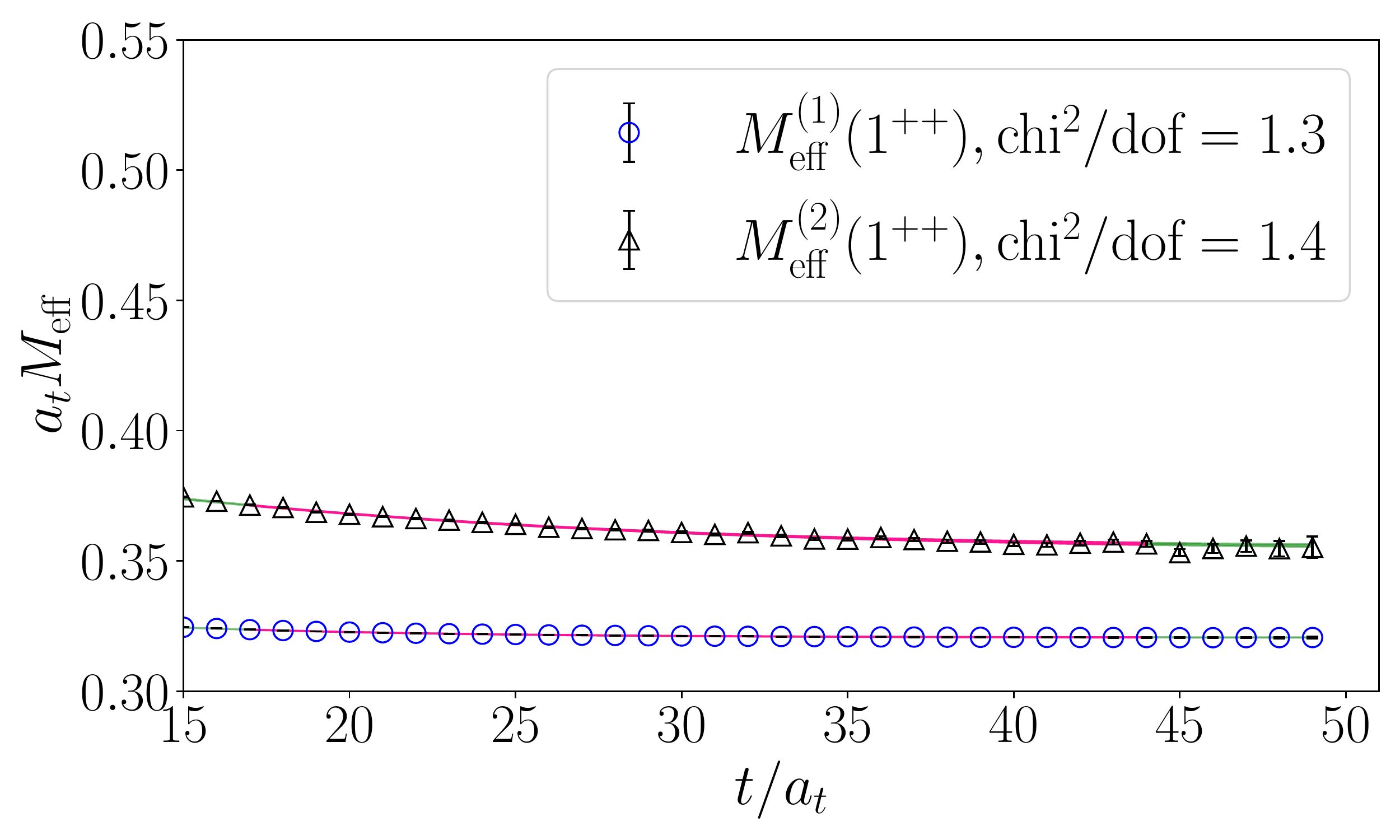}
	\includegraphics[height=3.5cm]{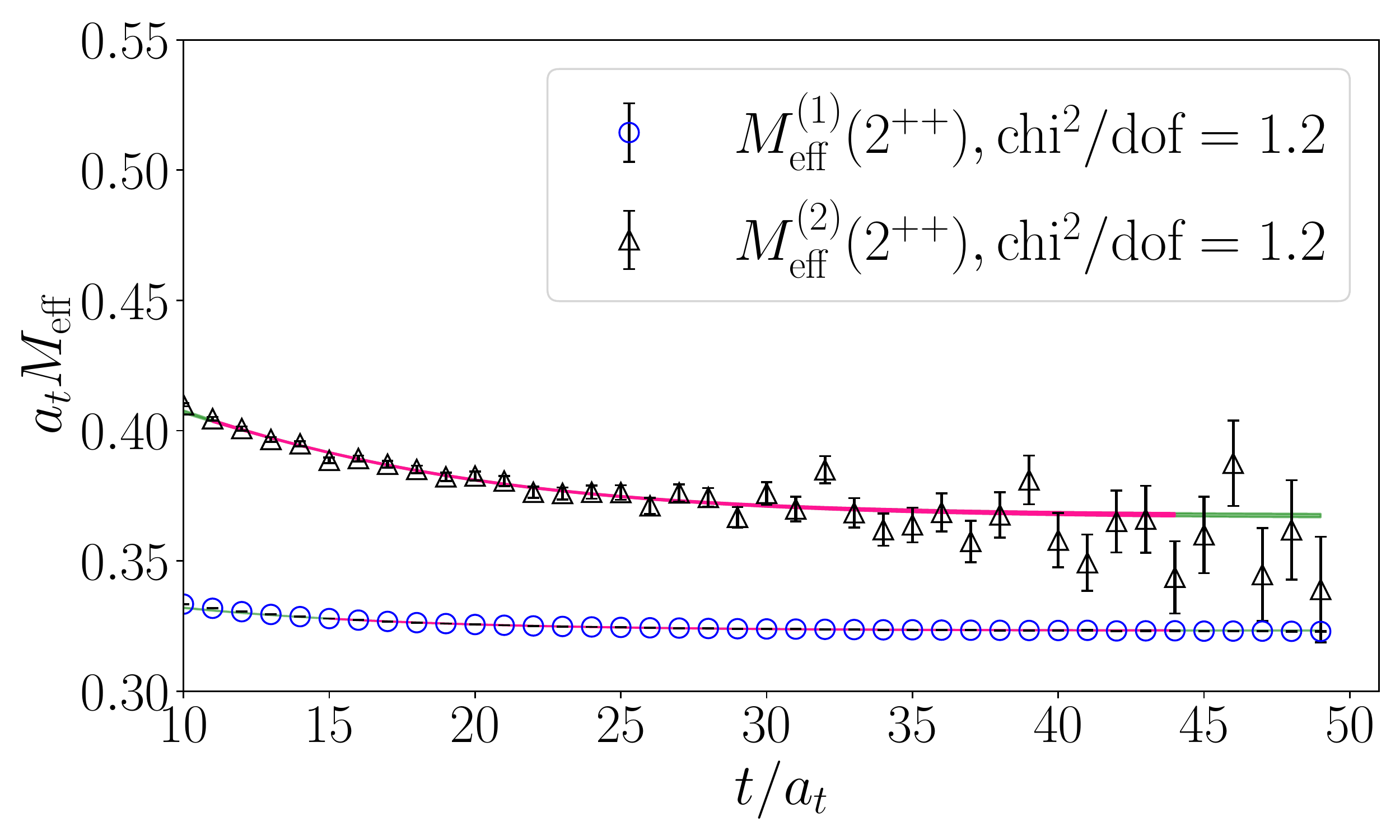}
	\includegraphics[height=3.5cm]{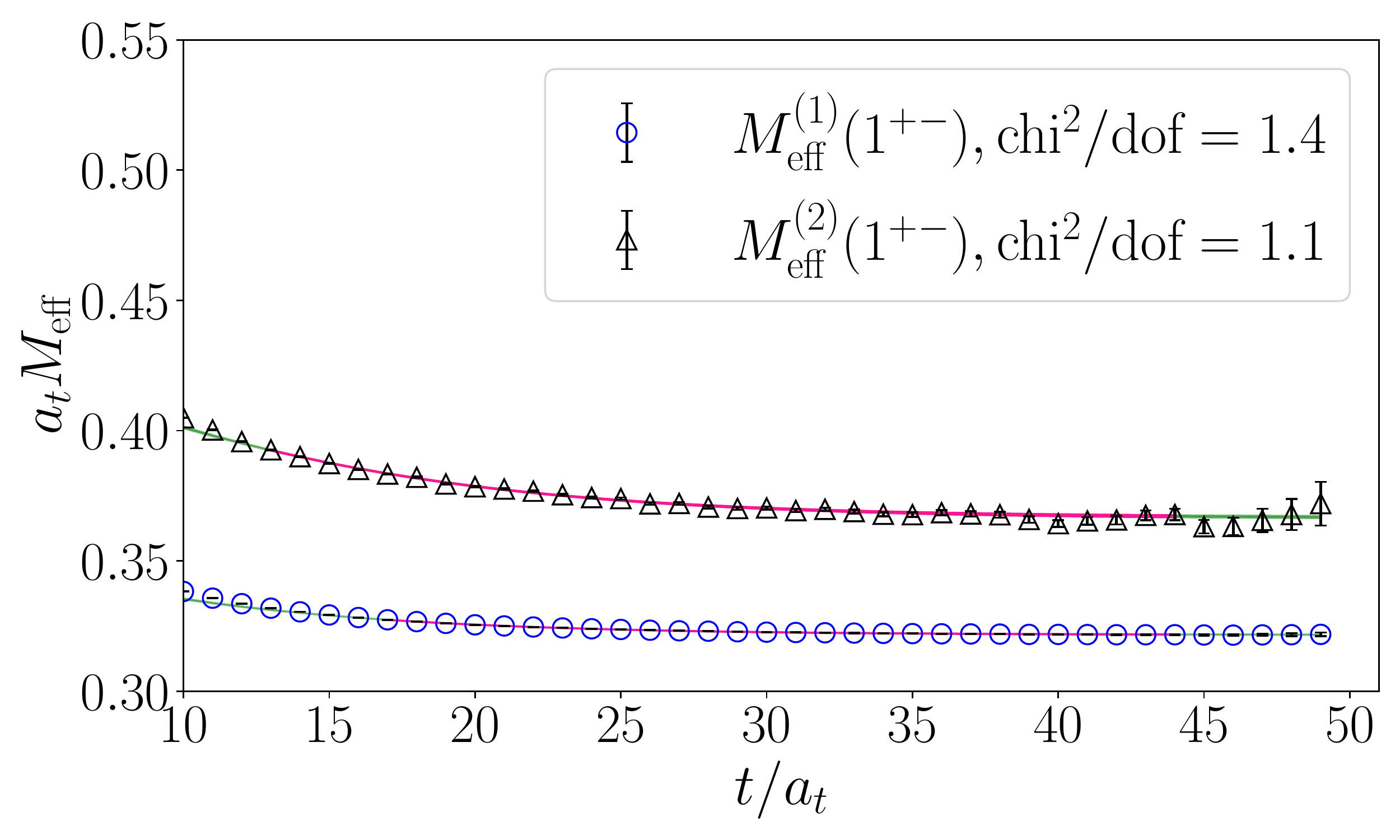}
	\caption{\label{fig:varE1} The mass plateaus of the correlation functions of the optimized operators obtained through GEVP method on gauge ensemble E1. In each of the six channels $J^{PC}=0^{-+},1^{--},(0,1,2)^{++}$ and $1^{-+}$, the mass plateaus corresponding to the lowest two states are shown. For each correlation function, a two-mass-term fit 
		is performed with the second mass term being introduced to take into account the residual contamination of higher states. The darker colored bands illustrate the fit results using the corresponding time window.}
\end{figure*}

\begin{figure*}[t!]
	\includegraphics[height=3.5cm]{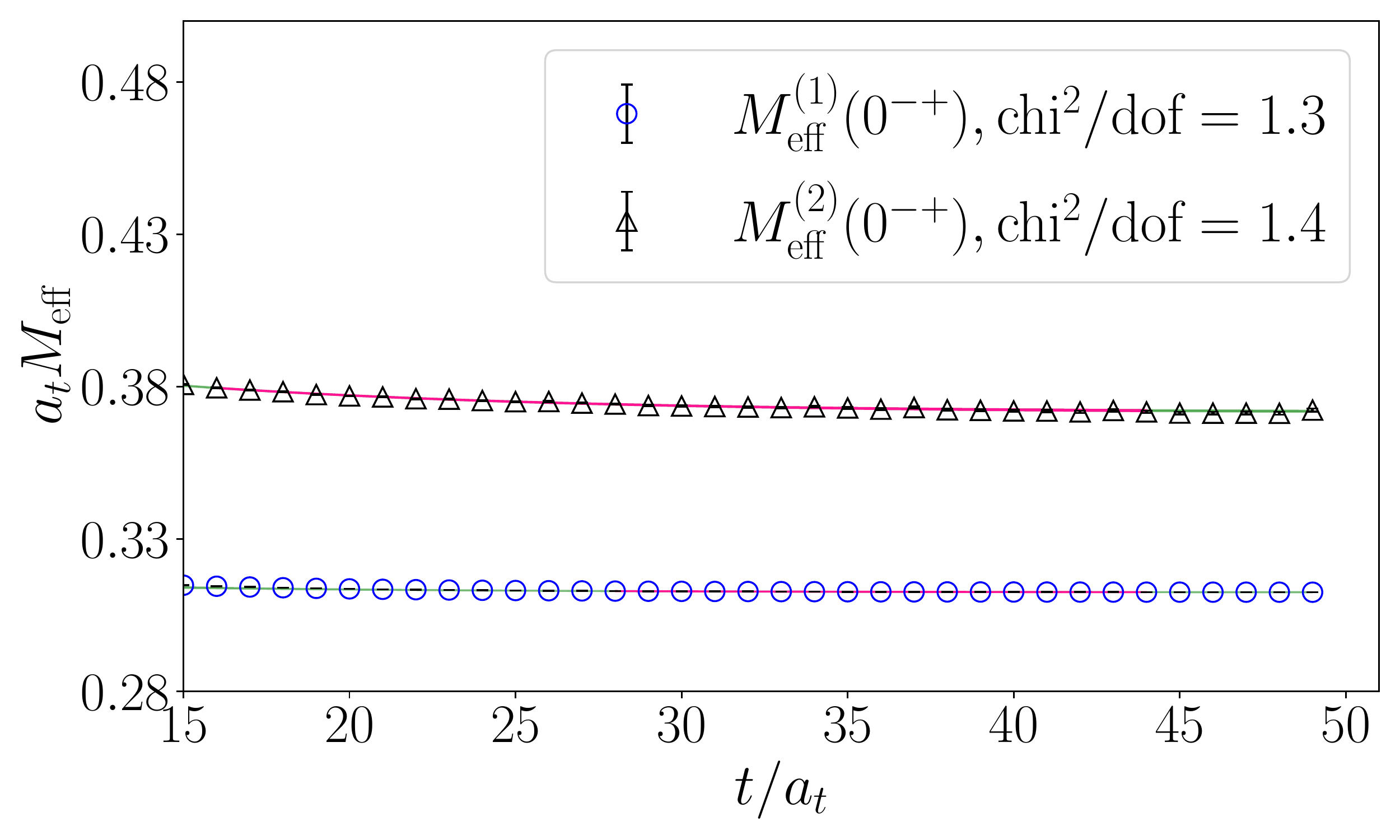}
	\includegraphics[height=3.5cm]{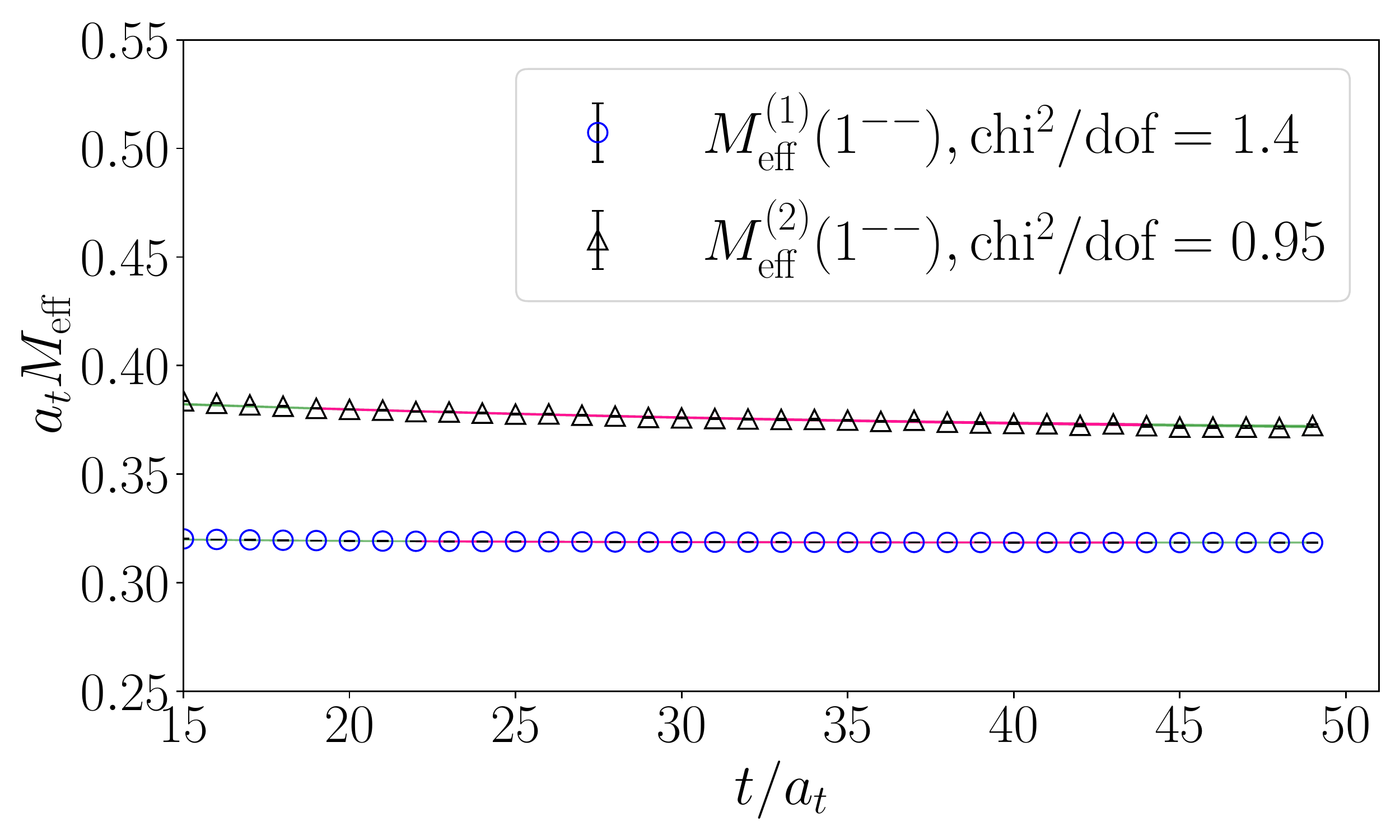}
	\includegraphics[height=3.5cm]{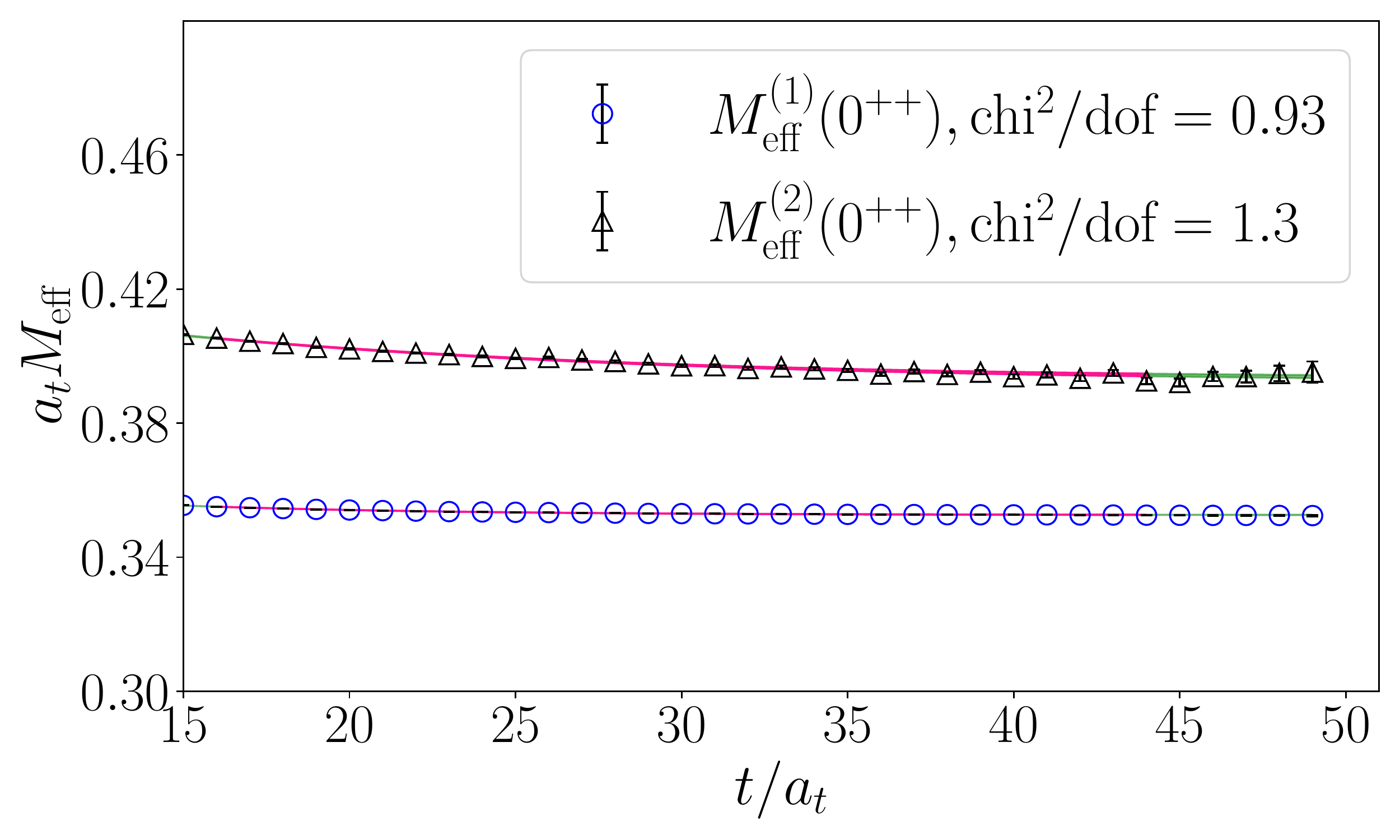}
	\includegraphics[height=3.5cm]{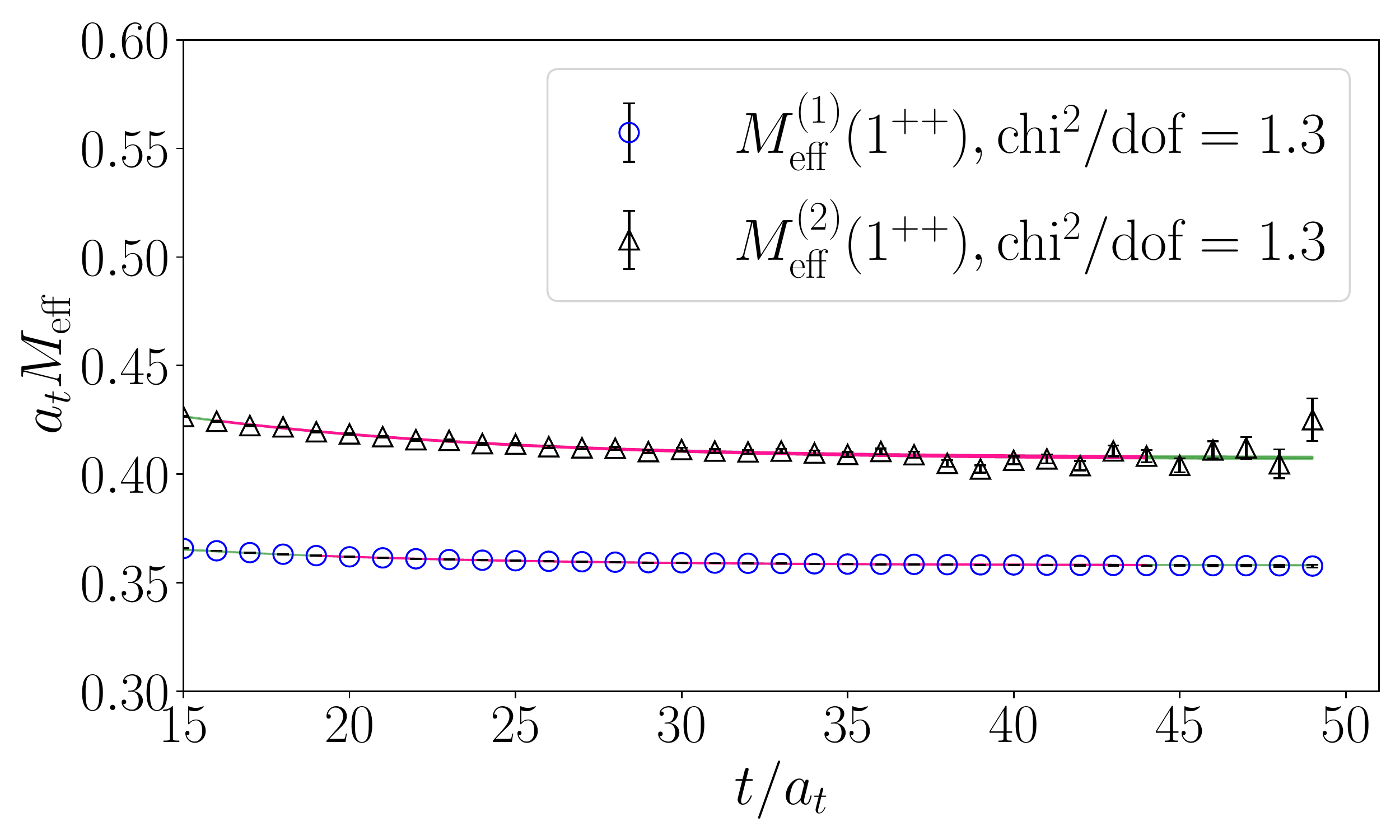}
	\includegraphics[height=3.5cm]{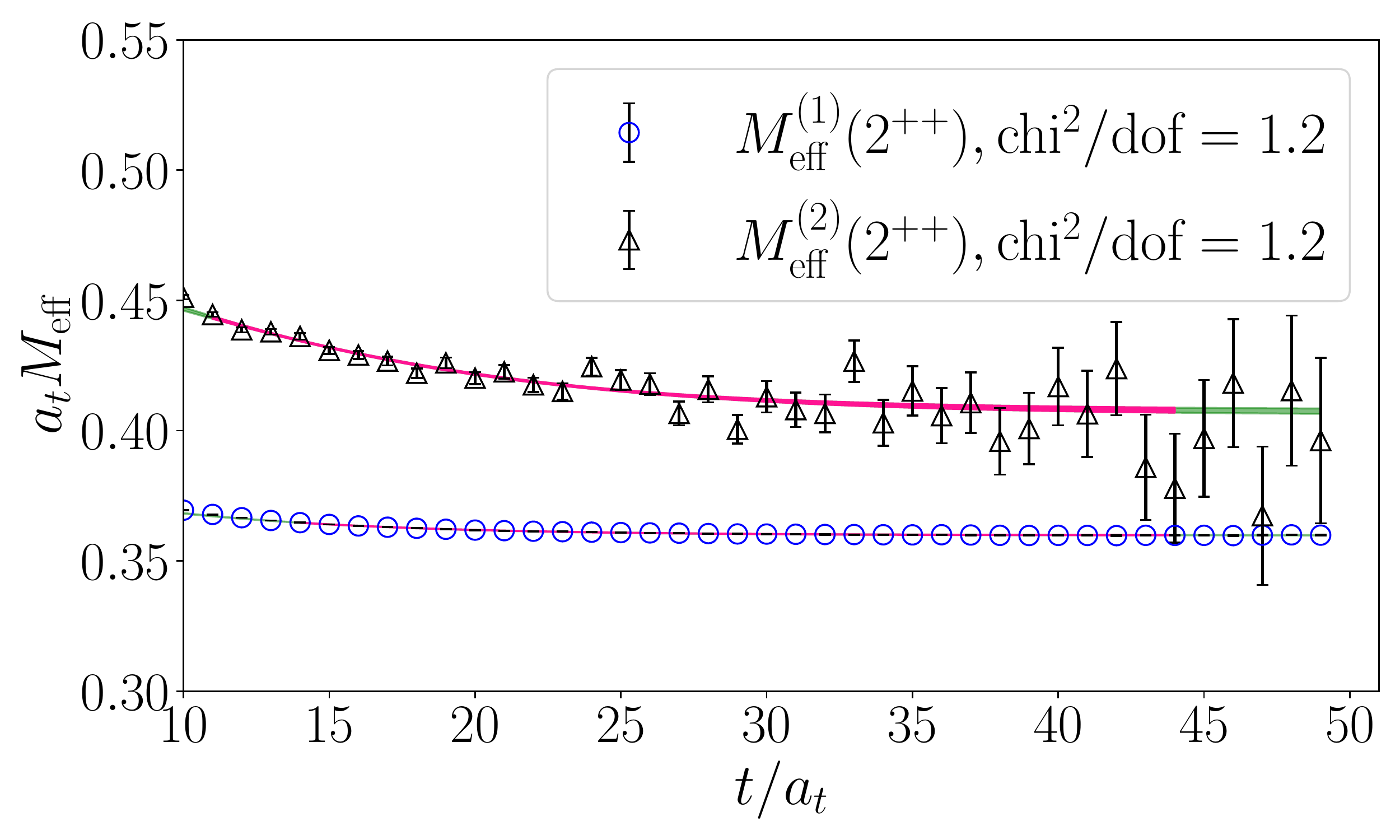}
	\includegraphics[height=3.5cm]{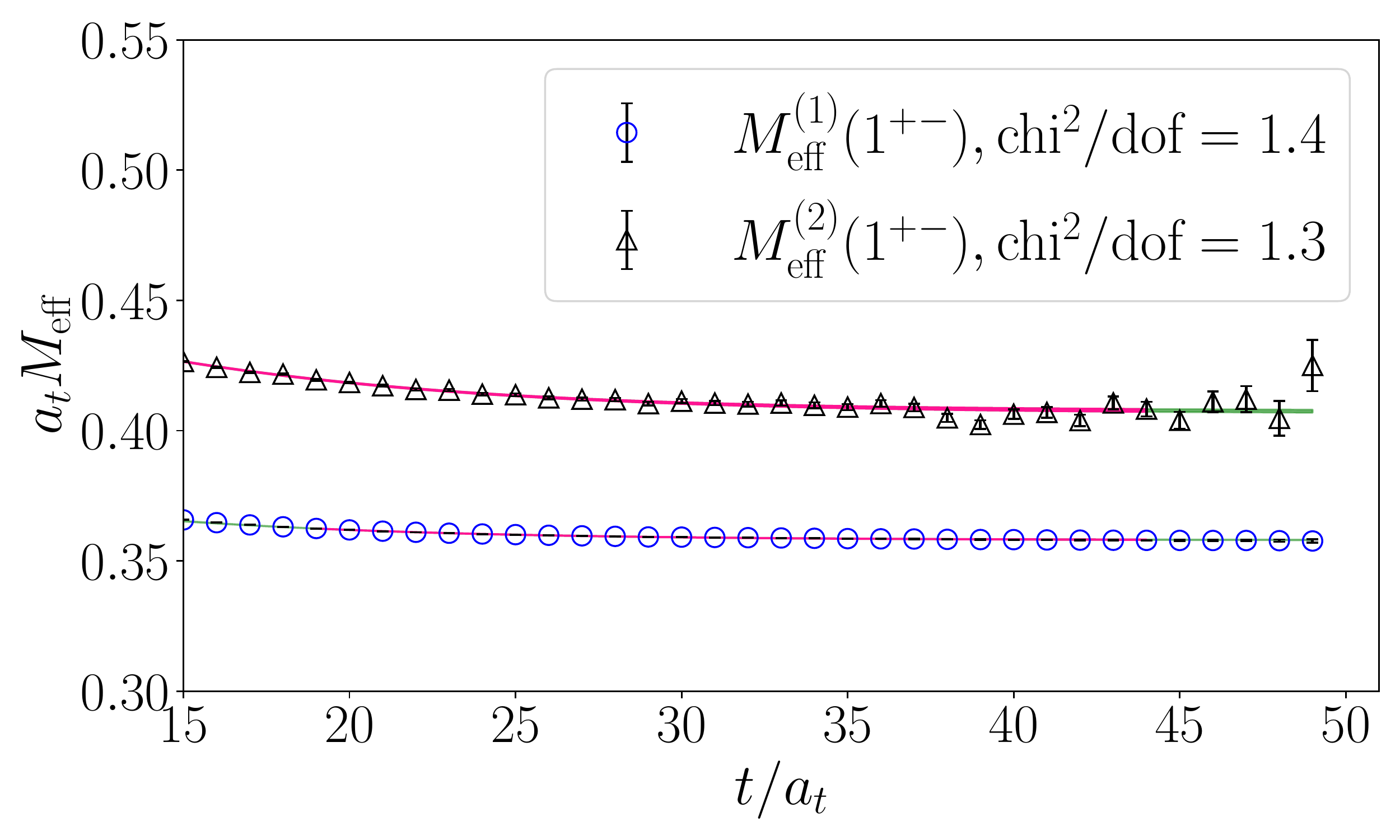}
	\caption{\label{fig:varE2} The mass plateaus of the correlation functions of the optimized operators obtained through GEVP method on gauge ensemble E2. In each of the six channels $J^{PC}=0^{-+},1^{--},(0,1,2)^{++}$ and $1^{-+}$, the mass plateaus corresponding to the lowest two states are shown. For each correlation function, a two-mass-term fit 
		is performed with the second mass term being introduced to take into account the residual contamination of higher states. The darker colored bands illustrate the fit results using the corresponding time window.}
\end{figure*}

In practice, $\mathbf{r}$ are chosen to be spatially on-axis displacements with $|\mathbf{r}|/a_s=0,1,2,3,4,5,6,7$. The differences between the different operators can be monitored through the effective masses of the corresponding correlation functions $C(t)$ of the operators
\begin{equation}\label{eq:eff-m}
a_t M_\mathrm{eff}(t)=\ln \frac{C(t)}{C(t+1)}.
\end{equation}
$a_t M_\mathrm{eff}(t)$ of these operators in the $0^{-+}$ channel are shown in Fig.~\ref{fig:VarR}. The different $t$-behaviors of the effective mass plateaus 
show the different coupling of the operators with different $|\mathbf{r}|$ to different intermediate states. For a given quantum number $J^{PC}$, the differences among different operators $O_i$ in an operator set $\{O_i, i=1,2,\ldots\}$ can be reflected by their couplings to the orthogonal complete state set $\{|n\rangle, n=1,2,\ldots\}$ with the normalization condition $\langle n|m\rangle =\delta_{mn}$,
\begin{equation}
O_i^\dagger|0\rangle\equiv \sum\limits_{n} |n\rangle\langle n|O_i^\dagger|0\rangle\equiv\sum\limits_n W_{in}^*|n\rangle
\end{equation}
where $W_{in}=\langle 0|O_i|n\rangle$ is defined. Therefore, the differences of $W_{in}$ show how different the operators in this operator set are, and also result in the different $t$-behaviors of the effective masses $M_\mathrm{eff}(t)$ of the corresponding correlation functions in the early time region. Since in each channel we only focus on the lowest two states, we select operators $\mathcal{O}(r,t)$ with $r/a_s=0,3,6$ to compose an operator set, which are expected to couple to the intermediate states more differently, as manifested by the effective masses of these correlation functions. Based on the operator set in each channel, we first carry out the variational method analysis to obtain the optimized operators that couple most to specific states. That is to say, we calculated the correlation
matrix $\{C_{ij}(t)\}$ of each operator set and solve the generalized eigenvalue problem 
\begin{equation}
C_{ij}(t)v_j=\lambda(t,t_0)C_{ij}(t_0)v_j
\end{equation} 
to get the eigenvector $v_i^{(n)}$ that corresponds to the $n$-th largest eigenvalue $\lambda^{(n)}(t,t_0)$. It is expected that the operator $\mathcal{O}^{(n)}(t)=v_i^{(n)}\mathcal{O}(r_i,t)$ couples most with the $n$-th lowest state. Thus we have the correlation function $C^{(n)}(t)$ of the optimized operator $\mathcal{O}^{(n)}$,
\begin{equation}
C^{(n)}(t)=\langle \mathcal{O}^{(n)}(t)\mathcal{O}^{(n),\dagger}(0)\rangle\equiv v_i^{(n)}v_j^{(n)}C_{ij}(t),
\end{equation}
whose effective mass function $a_t M_\mathrm{eff}^{(n)}(t)$ can be defined similarly to Eq.~(\ref{eq:eff-m}). $a_t M_\mathrm{eff}^{(n)}(t)$ functions for $n=1,2$ for all the channels are plotted in Fig.~\ref{fig:varE1} (ensemble E1) and Fig.~\ref{fig:varE2} (ensemble E2). $a_t M_\mathrm{eff}^{(1)}(t)$ and  $a_t M_\mathrm{eff}^{(2)}(t)$ are clearly separated from each other, but still have mild time dependence in the early time range due to slight contamination from higher states. In order to obtain the mass values more precisely, we perform two-mass-term fits to $C^{(n)}(t)$
\begin{eqnarray}
C^{(n)}(t)&=&W_1 \left(e^{-M_1 t}+e^{-M_1 (T-t)}\right)\nonumber\\
&+&W_2 \left(e^{-M_2 t}+e^{-M_2 (T-t)}\right).
\end{eqnarray}
where the second mass term is adopt to account for the higher state contamination. $W_1$ and $W_2$ are coefficients of the first mass term and the second mass term respectively.  
The fit results with the fitted parameters are also plotted in Fig.~\ref{fig:varE1} and Fig.~\ref{fig:varE2} as curves with error bands, where the extensions of the curves illustrate the fit windows. For all the fits, the values of the $\chi^2$ per degree of freedom ($\chi^2/\mathrm{dof}$) are around one and exhibit the fit quality. The fitted masses of $1P$, $2P$ charmonium states are tabulated in Table~\ref{tab:nPmass}, in which the experimental values are also given for comparison.

\begin{table*}[t]
    \centering \caption{\label{tab:nPmass} The masses of 1P and 2P states derived from ensemble E1 and E2. The masses are converted into the values in physical units. The 'center of gravity' masses are the spin averages of the spin-triplet $nP$ charmonium masses. The experimental results are also presented for comparison.}
    \begin{ruledtabular}
        \begin{tabular}{ccccccc}
            &$J^{PC}$   & $\chi_{c0}^{(')}(0^{++})$ & $\chi_{c1}^{(')}(1^{++})$    &$\chi_{c2}^{(')}(2^{++})$    & $h_c^{(')}(1^{+-})$ & $m_\mathrm{COG}$\\
            E1  & $M(1P)(\mathrm{MeV})$ & 3039.1(4) & 3083.8(6) & 3109.7(6)  & 3094.0(7) & 3093.2(4) \\
            & $M(2P)(\mathrm{MeV})$ & 3410(4) & 3412(7)   & 3531(8) &  3526(5) & 3478(5) \\
            &&&&&&\\
            E2  &  $M(1P)(\mathrm{MeV})$  &  3391.8(6) & 3433.8(9)    & 3461.8(7)  & 3443.8(9)  & 3444.7(5) \\
            &  $M(2P)(\mathrm{MeV})$  & 3785(6)  & 3792(9)    & 3916(11) & 3917(6)  & 3860(6) \\
            &&&&&&\\
            Exp.& $M(1P)(\mathrm{MeV})$  & 3414.7(3) & 3510.7(1) & 3556.2(1) & 3525.4(1) & 3525.31(7) \\
                & $M(2P)(\mathrm{MeV})$ & 3862(51) & 3871.7(1) & 3922(1) & &3898(6) \\
            &&&&&&
        \end{tabular}
    \end{ruledtabular}
\end{table*}

The non-relativistic quark model expects that for the same principal quantum number $n$, the $n^{2S+1}L_J$ multiplet with $S=1$ and $L\neq 0$ has a 'center of gravity' mass $M_\mathrm{COG}$, which is the spin average mass of this multiplet and should be equal to the mass of the spin singlet counterpart (the derivation of the relation can be found in the Appendix A). For example, for the $L=1$ multiplets $n^3P_J$ multiplet, $M_\mathrm{COG}$ is defined as  
\begin{eqnarray}\label{eq:COG}
M_\mathrm{COG}(nP)&=&\frac{1}{9}\left[5M(n^3P_2)+3M(n^3P_1)+M(n^3P_0)\right],\nonumber\\
&=&M(n^1P_1).
\end{eqnarray} 
As shown in Table~\ref{tab:nPmass}, the masses of the $1P$ states are determined very precisely with the statistic error being less than 1 MeV, so we can check the relation of Eq.~(\ref{eq:COG}). The difference between the $M_\mathrm{COG}(nP)$ and the mass of the spin singlet $h_c(nP)$ is sometimes called the hyperfine splitting of $nP$ charmonium states, which is denoted by $\Delta_\mathrm{HFS}(nP)=\Delta_\mathrm{HFS}=M_{h_c}(nP)-M_\mathrm{COG}(nP)$ in this paper. On the two ensembles (E1 and E2) in this paper, we obtain $\Delta_\mathrm{HFS}(1P)$ as 
\begin{eqnarray}
E1:~~\Delta_\mathrm{HFS}(1P)&=&~~0.8\pm 0.8~\mathrm{MeV}\nonumber\\
E2:~~\Delta_\mathrm{HFS}(1P)&=&-0.9\pm 1.0~\mathrm{MeV}.
\end{eqnarray}
In other words, the relation Eq.~(\ref{eq:COG}) is satisfied for $1P$ states with a high precision. For $2P$ states, we have 
\begin{eqnarray}
E1:~~\Delta_\mathrm{HFS}(2P)&=&49\pm 7~\mathrm{MeV}\nonumber\\
E2:~~\Delta_\mathrm{HFS}(2P)&=&57\pm 9~\mathrm{MeV}.
\end{eqnarray}
There is a substantial deviation from Eq.~(\ref{eq:COG}). We are not sure the reason for this deviation yet. It is possible that $2P$ charmonium states do have a non-zero hyperfine splitting. It is also possible that the masses of $2P$ states are not determined precisely enough. This should be explored in future studies. 

It is interesting to note that the experimental results support the relation of Eq.~(\ref{eq:COG}) to a very high precision~\cite{Zyla:2020zbs}. For charmonium systems, the $1P$ states $(h_c(1^1P_1),\chi_{c0,1,2}(1^3P_{0,1,2}))$ have been well established. According to the PDG data, the 'center of gravity' mass of $1P$ charmonia is $M_\mathrm{COG}=3525.3(1)$ MeV, which is almost the same as the $h_c$ mass $M(h_c)=3525.4(1)$ MeV. For bottomium systems, the $1P$ and $2P$ states are below the $B\bar{B}$ threshold and have very small widths. They are approximately stable particles and have direct correspondence to the according states predicted by the non-relativistic quark model. The $M_\mathrm{COG}(1P)=9899.9(5)$ MeV and $M_\mathrm{COG}(2P)=10260.3(6)$ also reporduce the $h_b(1P)$ mass $M(h_b(1P))=9899.3(8)$ MeV and the $h_b(2P)$ mass $M(h_b(2P))=10259.8(1.2)$ MeV. 

If the relation described by Eq.~(\ref{eq:COG}) is somewhat universal, we can use it to make a prediction of the mass of $h_c(2P)$. Experimentally, there are three particles being assigned to be the candidates of the $\chi_{c0,1,2}$ charmonia, which are $\chi_{c1}'(3872)$
(the old $X(3872)$), $\chi_{c2}'(3930)$ and $\chi_{c0}'(3860)$, while there is no experimental evidence of $h_c'$ yet. The resonance parameters of these three states are 
\begin{eqnarray}
\chi_{c0}'(3860):&& M_R=3860(50)\mathrm{MeV},  \Gamma_R\sim 200~~\mathrm{MeV}\nonumber\\
\chi_{c1}'(3872):&& M_R=3871.7(2)\mathrm{MeV},  \Gamma_R<1.2~~\mathrm{MeV}\nonumber\\
\chi_{c2}'(3930):&& M_R=3922(1)\mathrm{MeV},  \Gamma_R=35(3)~~\mathrm{MeV}.
\end{eqnarray} 
Except that $\chi_{c0}'(3860)$'s width is large, those of the other two states are relatively small. Using Eq.~(\ref{eq:COG}), the mass of $h_c(2P)$ can be tentatively estimated to be 
\begin{equation}
M(h_c(2P))=M_\mathrm{COG}(2P)\approx 3898(6) ~\mathrm{MeV}.
\end{equation}

\par

\section{Some caveats}\label{sec:discussion}
We would like to point out the possible caveats of our lattice setup.
As we addressed in Sec.~\ref{sec:numerical}, the tadpole improved tree level clover action proposed by the Hadron Spectrum Collaboration is adopted for charm quarks in this study~\cite{Edwards:2008ja,HadronSpectrum:2008xlg}. Our calculation shows that the hyperfine splitting $\Delta_\mathrm{HFS}=M_{J/\psi}-M_{\eta_c}$ is approximately 60 MeV for the two ensembles E1 and E2. This is obviously very far from the experimental value $\Delta_\mathbf{HFS}=113(3)$ MeV, and signals large discretization errors. The Hadron Spectrum Collaboration also found this large discrepancy in its calculation of the charmonium spectrum~\cite{HadronSpectrum:2012gic}, where $\Delta_\mathrm{HFS}=80(2)$ MeV was obtained. This discrepancy might be attributed to clover term $\bar{\psi}\sigma_{\mu\nu}F_{\mu\nu}\psi$ of the fermion action, which results in the spin-orbital and spin-spin interactions between the charm quark and antiquark in the non-relativistic approximation. It is found~\cite{HadronSpectrum:2012gic} that a larger coeffecient of the spatial clover term $\sigma_{ss'}\hat{F}_{ss'}\propto \mathbf{\sigma}\cdot \mathbf{B}$ can give a larger $\Delta_\mathrm{HFS}$. On the other hand, even the masses of the $1P$ and $2P$ charmonium states are determined very precisely in this work (see in Table~\ref{tab:nPmass}), it is obvious that the fine splittings between the $1P$ states are also smaller than those of the experimental values. In the non-relativistic approximation, these splittings are due to the spin-orbital and the tensor interactions~\cite{Lucha:1991vn} (also in Appendix A). The under-determined splittings in this work may imply that the coefficient of the temporal clover term $\sigma_{ts}F_{ts}\propto \mathbf{\sigma}\cdot\mathbf{E}$ of the fermion action also has large discretization uncertainties. It may also because we ignored light u,d,s quarks. Based on these observations, it seems that the fermion action we adopt in this work may not be a very good anisotropic version for charm quarks. It is noted that another version of the clover fermion action on the anisotropic lattices~\cite{Liu:2001ss,Zhang:2001in} can produce a larger $\Delta_\mathrm{HFS}$~\cite{Yang:2012mya,Gui:2019dtm,Ma:2019hsm}. Of course, whatever the lattice setup is, the continuum limit should be the same, but the smaller the discretization uncertainties, the smaller systematic errors after the continuum extrapolation. 

\section{Summary}\label{sec:summary}
In this work, we generate gauge ensembles with two flavor degenerate dynamical charm quarks on anisotropic lattices. This lattice setup enables us to investigate the contribution of charm quark annihilation diagrams to charmonium masses in an unitary theoretical framework for charm quarks. The distillation method is adopted to realize both the smearing 
scheme of quark fields and the disconnected diagrams in calculating meson correlation functions. With large statistics, the effects of disconnected diagrams can be derived with a high precision. 

For a given quantum number $J^{PC}$, the mass shift of charmonium masses due to the disconnected diagrams can be derived from the ratio $R(t)$ of the disconnected part of the charmonium correlation function to the connected part. It is found that the $R(t)$ are relatively large for the scalar and the pseudoscalar channel, and the time dependence of $R(t)$ imply that there may be contribution from lighter states than the corresponding lowest charmonium states to the correlation functions when the disconnected diagrams are considered. This is not surprising since glueballs can appear as intermediate states in this case, whose masses are predicted by lattice QCD to be lower than charmonium states. By considering the contribution of glueballs, the mass shift of the ground state pseudoscalar charmonium $\eta_c$ is +3.0(1) MeV and +3.1(2) MeV, respectively,for the two gauge ensembles involved, which is consistent with the previous result in Ref.~\cite{Levkova:2010ft}. These values are also consistent with mass shift 3.9(9) MeV derived from the glueball-charmonium mixing mechanism~\cite{Zhang:2021xvl}. For the scalar charmonium $\chi_{c0}$, we cannot get a reliable result of the mass shift yet. The contribution of the charm quark annihilation effect to the $J/\psi$ mass is very tiny as expected by the OZI rule, since the two charm quark loops are mediated by at least three intermediate gluons. For $\chi_{c1}$ and $h_c$, the mass shifts due to the charm annihilation effect are approximately 1 MeV, which are smaller than that of $\eta_c$. This is understandable since the contribution of two-gluon intermediate states is suppressed to some extent due to the Landau-Yang theorem~\cite{Landau:1948kw,Yang:1950rg}. It is surprising that this kind of mass shift of $\chi_{c2}$ is relatively large and negative (-3.0(3) MeV for E1 and -2.8(2) for E2). We don't have  an explanation to this yet. Lattice QCD studies predict that the lowest tensor glueball mass is approximately 2.2-2.4 GeV~\cite{Morningstar:1999rf,Chen:2005mg,Sun:2017ipk,Richards:2010ck,Gregory:2012hu}, which is lower than the $\chi_{c2}$ mass and should results in positive mass shift if $\chi_{c2}$ and the tensor glueball can mix. Note that in this study we only consider the charm sea quark effects, the effects of $u,d,s$ quarks have not been taken into consideration yet, which are expected theoretically to affect the mass spectrum of charmonia also~\cite{Hatton:2020qhk}. In the presence of $u,d,s$ quarks, the situation will be much more complicated due to charmoniums decays and appearance of light hadron intermediate states. Therefore, the study of their impact on charmonium masses is still a challenging task in the present stage of lattice QCD. 

The relation $M_\mathrm{COG}(nP)=M_{h_c}(nP)$ is expected by the non-relativistic quark model. We observe that this relation is satisfied to a very high precision level for $1P$ charmonia, namely $M_\mathrm{COG}(1P)-M_{h_c}(1P)=0.8\pm 0.8$ MeV and $0.9\pm 1.0$ MeV at the two charm quark masses used in this work. For the $2P$ charmonia, we observe a large deviation from this relation. There are two possible reasons for this deviation: this relation actually does not hold for $2P$ states, or the masses of the $2P$ states derived in this work have contaminations from higher states. This should be explored in future studies. On the other hand, for the two charm quark masses, the mass splittings between the $1P$ and $2P$ charmonium we obtained are smaller than those of the non-relativistic quark model predictions~\cite{Godfrey:1985xj,Barnes:2005pb}, but consistent with experimental results, given the assignment that $\chi'_{c0}(3860)$, $\chi_{c1}'(3872)$ and $\chi'_{c2}(3930)$ are $2P$ charmonium states.  

\section*{acknowlegement}
 This work is supported by the Strategic Priority Research Program of Chinese Academy of Sciences (No. XDB34030302), the National Key Research and Development Program of China (No. 2020YFA0406400), and the National Natural Science Foundation of China (NNSFC) under Grants No.11935017, No.11775229, No.12075253, No.12070131001 (CRC 110 by DFG and NNSFC). 
 Y. Chen also acknowledges the support of the CAS Center for Excellence in Particle Physics (CCEPP). The Chroma software system~\cite{Edwards:2004sx} and QUDA library~\cite{Clark:2009wm,Babich:2011np} are acknowledged.  The computations were performed on the CAS Xiandao-1 computing environment, the HPC clusters at Institute of High Energy Physics (Beijing) and China Spallation Neutron Source (Dongguan), and the GPU cluster at Hunan Normal University.
\begin{table}[t]
	\centering \caption{\label{tab:quantum-number} The expectation values of $\langle \mathbf{L}\cdot\mathbf{S}\rangle$ and $\langle S_{12}\rangle$ for $S=1$ and $L\neq 0$ states. }
	\begin{ruledtabular}
		\begin{tabular}{cccc}
			$J$    &  $L-1$   & $L$   & $L+1$ \\\hline
			$\langle \mathbf{L}\cdot\mathbf{S}\rangle_J$    
			&  $-(L+1)$ & $-1$ &   $L$ \\
			$\langle S_{12}\rangle_J$   
			&  $-\frac{2(L+1)}{2L-1}$   & $2$   &  $-\frac{2L}{2L+3}$\\
		\end{tabular}
	\end{ruledtabular}
\end{table}
\section*{Appendix }
\subsection*{A. `Center of gravity' mass of heavy quarkonia}
In the quark model picture, the spectroscopy of charmonia can be studied 
by solving the non-relativistic Schr\"{o}dinger equation of bound states with QCD-inspired inter-quark potentials $V(r)$ along with the relativistic corrections. To order $v^2\approx \mathbf{p}^2/m^2$, where $v$ is the velocity of the (anti-)charm quark with mass $m$ within a bound state, the generalized Breit-Fermi Hamiltonian~\cite{Lucha:1991vn} is 
\begin{eqnarray}
H&=&2m+\frac{\mathbf{p}^2}{m}-\frac{1}{4m^3}\mathbf{p}^4+V(r)\nonumber\\
&&+H_\mathrm{SI}+H_\mathrm{LS}+H_\mathrm{SS}+H_\mathbf{T}
\end{eqnarray}
where $H_\mathbf{SI}$ is the spin-independent interaction term whose explicit
form is irrelevant of the discussion here and is omitted, $H_\mathbf{LS}$ is the term for the spin-obital coupling, $H_\mathrm{SS}$ is the spin-spin interaction term, and $H_\mathrm{T}$ is the tensor interaction part. If the potential $V(r)$ can be splitted into the vector-like part $V_\mathrm{V}(r)$ 
and scalar-like part $V_\mathrm{S}(r)$, the explicit expressions of $H_\mathrm{LS}$, $H_\mathrm{SS}$ and $H_\mathrm{T}$ are 
\begin{eqnarray}
H_\mathrm{LS}&=&\frac{1}{2m^2 r}\left(3 \frac{d}{dr}V_\mathrm{V}-\frac{d}{dr}V_\mathrm{S}\right)\mathbf{L}\cdot\mathbf{S},\nonumber\\
H_\mathrm{SS}&=&\frac{2}{3m^2}\mathbf{S}_1\cdot\mathbf{S}_2 \nabla^2  V_\mathrm{V}(r),\nonumber\\
H_\mathrm{T}&=& \frac{1}{12m^2}\left(\frac{1}{r} \frac{d}{dr} V_\mathrm{V}-\frac{d^2}{dr^2} V_\mathrm{V} \right)S_{12}
\end{eqnarray}
with 
\begin{equation}
S_{12}\equiv 12\left(\frac{(\mathbf{S}_1\cdot\mathbf{r})(\mathbf{S}_2\cdot\mathbf{r})}{r^2}-\frac{1}{3}\mathbf{S}_1\cdot\mathbf{S}_2 \right)
\end{equation}
where $\mathbf{S}_{1,2}$ are the spins of the charm quark and antiquark, $\mathbf{S}=\mathbf{S}_1+\mathbf{S}_2$ is their total spin. 
For a given state in terms of the eigenvalues $S,L,J$ of $\mathbf{S}$, the orbital momentum $\mathbf{L}$ and the total angular momentum $\mathbf{J}=\mathbf{L}+\mathbf{S}$, the expectation value $\langle \mathbf{L}\cdot\mathbf{S}\rangle$ reads
\begin{equation}
\langle \mathbf{L}\cdot\mathbf{S}\rangle=\frac{1}{2}\left[J(J+1)-L(L+1)-S(S+1)\right].
\end{equation}
Obviously this expectation value for $L=0$ or $S=0$ states vanishes. It is easy to see that the expectation value of $S_{12}$ also vanishes for $L=0$ or $S=0$. Otherwise the expectation value of $S_{12}$ reads
\begin{equation}
\langle S_{12}\rangle = \frac{4}{(2L+3)(2L-1)}\left[\langle \mathbf{S}^2 \mathbf{L}^2\rangle-\frac{3}{2}\langle \mathbf{L}\cdot\mathbf{S}\rangle-3\langle \mathbf{L}\cdot\mathbf{S}\rangle^2\right].
\end{equation} 
As for the spin-spin interaction, if the $V_\mathrm{V}(r)$ is taken to be 
Coulomb-like, namely, $V_\mathrm{V}\propto 1/r$, then $H_\mathrm{SS}\propto \delta^3(\mathbf{r}) \mathbf{S}_1\cdot\mathbf{S}_2$ is actually a contact interaction, whose expectation value vanishes for $L\neq 0$ states, since their radial wave function $\phi_{nl}(r)$ at the origin $r=0$ is zero. For $L=0$ states ($n^{1}S_0$ and $n^{3}S_1$ states), the spin-spin interaction 
results the hyperfine splitting $\Delta M_\mathrm{hfs}$ between the spin triplet and spin singlet $S$ states. 

The values of $\langle \mathbf{L}\cdot\mathbf{S}\rangle$ and $\langle S_{12}\rangle$ for $S=1$ and $L\neq 0$ states are listed in Table~\ref{tab:quantum-number}. It is easy to see that for a $n^{3}L_J$ super-multiplet with $J=L-1,L,L+1$, one has
\begin{eqnarray}
\langle \mathbf{L}\cdot\mathbf{S}\rangle_\mathrm{avg}
&\equiv& \frac{1}{3(2L+1)}\sum\limits_J (2J+1)\langle\mathbf{L}\cdot\mathbf{S}\rangle_J =0 \nonumber\\
\langle S_{12}\rangle_\mathrm{avg}
&\equiv& \frac{1}{3(2L+1)}\sum\limits_J
(2J+1)\langle S_{12}\rangle_J =0.
\end{eqnarray}
Therefore, if one introduces a `center of gravity' mass $M_\mathrm{COG}(nL)$ for $n^{3}L_J$ states, 
\begin{equation}
M_\mathrm{COG}\equiv \frac{1}{2(2L+1)}\sum\limits_J (2J+1)M_J(nL),
\end{equation}
which is the spin averaged mass weighted by the number of polarizations, one should have 
\begin{equation}
M_\mathrm{COG}=M_L(n^1L_L),
\end{equation}
where $M_L(n^1L_L)$ is the mass of the spin singlet state of $nL$ supermultiplet.

\bibliography{ref}
\end{document}